\documentclass[aps,pra,preprint,amsmath,amsfonts,twoside]{revtex4-1}

\usepackage{graphicx}
\usepackage{xspace}
\usepackage{braket}
\usepackage{mathtools}

\newcommand{\figref}[1]{Fig.~\ref{#1}\xspace}
\newcommand{\eqab}[1]{Eq.~\eqref{#1}\xspace}
\newcommand{\rmat}{\textsl{R} matrix\xspace}
\newcommand{\rmath}{\textsl{R}-matrix\xspace}
\newcommand{\cora}{$A$\xspace}
\newcommand{\corb}{$B$\xspace}
\newcommand{\sche}{Schr\"odinger equation\xspace}
\newcommand{\abini}{\textit{ab initio}\xspace}
\newcommand{\bspl}{$B$-spline\xspace}
\newcommand{\bspls}{$B$-splines\xspace}
\newcommand{\gfn}{Green's function\xspace}

\begin{document}
	\title{Adiabatic potential energy curves of long-range Rydberg molecules: Two-electron R-matrix approach}
	\author{Michal \surname{Tarana}}
	\email{michal.tarana@jh-inst.cas.cz}
	\author{Roman \surname{\v{C}ur\'ik}}
	\affiliation{J. Heyrovsk\'y Institute of Physical Chemistry of the ASCR, v.v.i., Dolej\v{s}kova 2155/3, 182 23 Prague 8, Czech Republic}
	\begin{abstract}
		We introduce a computational method developed for study of long-range molecular Rydberg states of such systems that can be approximated by two electrons in a model potential of the atomic cores. Only diatomic molecules are considered. The method is based on a two-electron \rmath approach inside a sphere centered on one of the atoms. The wave function is then connected to a Coulomb region outside the sphere via multichannel version of the Coulomb \gfn. This approach is put into a test by its application to a study of Rydberg states of the hydrogen molecule for internuclear distances $R$ from 20 to 400 bohrs and energies corresponding to $n$ from 3 to 22. The results are compared with previous quantum chemical calculations (lower quantum numbers $n$) and computations based on contact potential models (higher quantum numbers $n$).
	\end{abstract}
	\maketitle
	\section{Introduction}
		\label{sec:introduction}
		Interactions of neutral atoms in their ground states with atoms in highly excited Rydberg states attracted considerable scientific attention (\cite{Beigman-rev,Omont1977} and references therein) since Fermi's  development of the theory explaining pressure shifts of atomic Rydberg spectral lines \cite{fermi1934}. \textcite{Greene-prl} discovered that the pair of atoms with one of them in the excited state and the other in the ground state (perturber), can form a long-range Rydberg molecules with large internuclear distances and unusual properties (trilobite states) \cite{Granger,Hamilton-PhD,Khuskivadze}. Later, \textcite{Hamilton2002} in their theoretical work predicted that another class of long-range Rydberg molecular states arises when the electron interaction with the perturber exhibits a shape resonance (butterfly states). These theoretical findings were experimentally verified \cite{Greene2006-prl,Bendkowsky2009} and they have been the focus of numerous theoretical \cite{Liu-Rost,Kurz,Junginger,Anderson} and experimental \cite{Vadla,Bendkowsky-prl,Bellos,Anderson-prl,Krupp-prl} investigations. It was found that the temperature and density of the alkali-metal atoms in a Bose-Einstein condensate are particularly favorable for the laser excitation of these long-range molecular Rydberg states.
		
		Initial theoretical papers \cite{Greene-prl,Granger,Hamilton2002} predicted the existence of the trilobite and butterfly states of Rb$_2$ at internuclear distances in the range $10^3-10^5$\thinspace a.u. with energies corresponding to $n\approx30$. Those works are based on an assumption that the interaction of the Rydberg electron with the neutral perturber can be approximated using the Fermi's pseudopotential \cite{fermi1934}.
		
		However, the measurements \cite{Greene2006-prl,Vadla,Bellos} were performed at lower energies corresponding to $n\approx 8\dots11$ and they show similar bound states of Rb$_2$ in this energy range. Calculations of the model potential energy curves (PECs) in this range of energies based on the Fermi's pseudopotential show local minima \cite{Greene-prl} that can be assigned to the measured spectral lines corresponding to the bound states. However, those minima of the PECs are located at internuclear distances $50-200$\thinspace a.u. Those values are considerably below the range $10^3-10^5$\thinspace a.u. for which the theory by \textcite{Greene-prl} was originally developed. Although the qualitative agreement of the theoretical models with the experimental spectra is encouraging, the quantitative validity of the Fermi's pseudopotential at shorter internuclear distances is not entirely clear. The purpose of this paper is to introduce a computational method that is capable of treating this range of smaller internuclear separations ($20 - 420$\thinspace a.u.) and lower energies corresponding to $n\approx8\dots20$ from the first principles. This will allow us to test the validity of Fermi's pseudopotential under these conditions.
		
		It is necessary to mention that more recent experimental works \cite{Bendkowsky-prl,Anderson-prl,Krupp-prl} focus on the energies corresponding to $n\approx35\dots46$. The measured results are in agreement with the theoretical model by \textcite{Greene-prl} applied to Rb$_2$ at internuclear separations $10^3-3\times10^3$\thinspace a.u.
		
		\textcite{Khuskivadze} developed an \abini method to calculate the PECs and dipole moments of the long-range excited molecular states. In this work the Rydberg atom is represented by the electron in the Coulomb potential, while the effects of the electronic structure of the atomic core are included in the calculation as a correction of the Coulomb wave functions parametrized by the quantum defects \cite{Davydkin}. The positively charged atomic core is placed at the distance $R$ from the neutral perturber that is represented by the short-range model potential in two lowest partial waves. The one-electron \sche with both potentials is solved in the sphere centered around the perturber and with a radius that allows the short-range model potential to vanish outside the sphere. The general solutions and their derivatives are then smoothly matched on the sphere with the wave function in the outer region. That matching is performed in terms of the Coulomb \gfn \cite{Hostler,Hostler1963}, since the only potential considered in the outer region is due to the positive charge of the Rydberg core. This technique guarantees that the correct bound-state asymptotic boundary conditions are satisfied and also that the complicated structure of the wave function in the outer region is treated analytically. In fact, it is necessary to evaluate the Coulomb \gfn and its radial derivative only on the matching sphere and the wave function elsewhere in the outer region does not explicitly appear in the calculation. Due to the bound-state boundary conditions, the wave function in the inner region can be smoothly matched with the wave function in the outer region only for discrete values of the energy that are identified as the adiabatic energies of the bound states. The method by \textcite{Khuskivadze} takes into account the finite size of the perturber as well as the variation of the Coulomb potential of the other positively charged atomic center throughout the region of the perturber. This effect becomes more relevant with decreasing internuclear separation.
		
		The two-electron computational method developed in this work is a generalization of the approach introduced by \textcite{Khuskivadze}. Particularly, the representation of the perturber and its interaction with the Rydberg atom is more advanced. Since the experimental studies dealing with the long-range molecular states are mainly focused on the systems involving alkali-metal atoms that can be well represented by single electron in the model potential, the approach developed in this paper is also based on that one-electron representation of the perturber.
		
		The interaction of the Rydberg electron with the valence electron of the perturber is a true two-electron repulsion. The additional degree of freedom due to the valence electron allows us to include the effect of the low excited states of the perturber in the theoretical model. It is particularly practical in the studies of systems, where the molecular Rydberg states associated with the second ionization threshold at energies below the first ionization threshold exist. Rb$_2$ is a good example of such system, since the energy of the dissociation channel $\mathrm{Rb}(5p)+\mathrm{Rb}(5p)$ is between the energies of the dissociation channels $\mathrm{Rb}(5s)+\mathrm{Rb}(6p)$ and $\mathrm{Rb}(5s)+\mathrm{Rb}(7p)$ \cite{Spiegelmann}.
		
		The two-electron formulation of the problem also allows us to treat the interaction of the perturber with the Rydberg electron more accurately. This is because the couplings between the Rydberg electron and the valence electron of the neutral perturber are in the present model treated in \abini manner. Only the closed-shell core of the perturber is represented by a model potential, while in the reference \cite{Khuskivadze} whole the neutral perturber was approximated by the model potential. That approximation required to introduce different model potentials for different spin states (singlet and triplet).
		
		The parameters of the model potential employed in \cite{Khuskivadze} to represent the perturber are optimized to reproduce its electron affinity along with the energies and widths of the low-energy atomic resonances. That interaction potential is restricted to two lowest partial waves (with respect to the center of the perturber), since the low-energy electron interaction with the neutral perturber in higher partial waves is predominated by the repulsive centrifugal barrier that prevents the electron from "probing" the details of the electronic structure of the perturber. However, the presence of another positively charged atomic core can reduce the strength of the repulsive centrifugal barrier and the Rydberg electron may be able to penetrate deeper into the perturber in higher partial waves.
		\begin{figure}
			\includegraphics{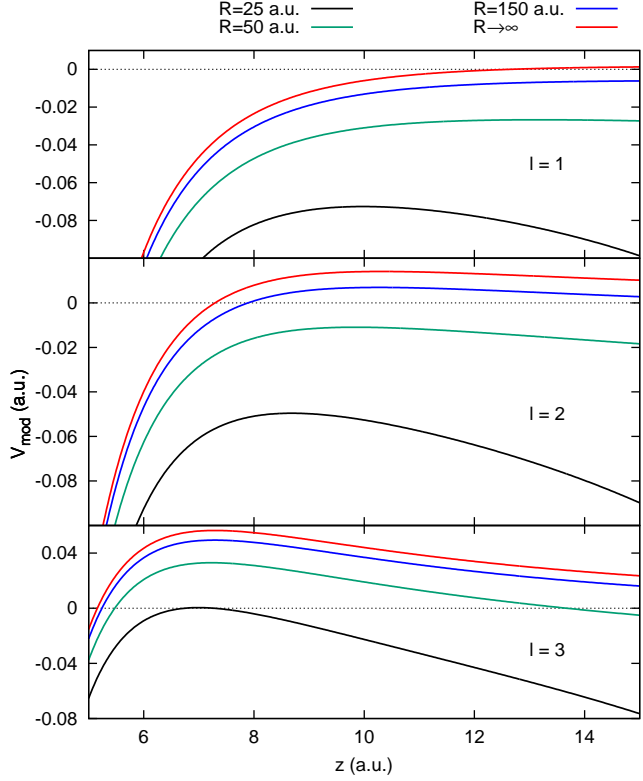}
			\caption{Illustration of the model potential of Rb$(5s)$ in the valence region of the perturber (its core is located at $z=0$) along the inernuclear axis $z$ calculated for different internuclear distances $R$ and for the partial waves $l=1,2,3$. The red curve denotes the potential in the absence of the other positively charged atomic core ($R\to\infty$).\label{fig:penetscheme}}
		\end{figure}
		This mechanism is illustrated in \figref{fig:penetscheme}. It shows the model potential $V_{\text{mod}}(z)$ of Rb$(5s)$ in the presence of another positively charged atomic core calculated along the internuclear axis for several values of the distance $R$:
		\begin{equation}
			V_{\text{mod}}(z)=
			\begin{dcases}
				V_{1S}(z)-\frac{1}{\left|z-R\right|}+\frac{l(l+1)}{2z^2}&l=1\vphantom{\frac{0}{0}}\\
				-\frac{\alpha_d}{z^4}-\frac{1}{|z-R|}+\frac{l(l+1)}{2z^2}&l=2,3,
			\end{dcases}
		\end{equation}
		where $V_{1S}(r=z)$ is a model potential taken from the reference \cite{Khuskivadze} [Eq. (17)], $\alpha_d=319.2$\thinspace a.u. is the static dipole polarizability of the rubidium atom \cite{Khuskivadze} and $l$ is the angular momentum with respect to the center of the perturber. For the purpose of this qualitative analysis it is sufficient to assume that the atomic potential beyond the core of Rb is predominated by the polarization potential. \figref{fig:penetscheme} shows that if the other positively charged core is sufficiently close, its Coulomb potential lowers the barrier for $l=1$ as well as for $l=2$ so that the electron in a higher Rydberg state can classically penetrate into the valence region. Therefore, the representation of the perturber in $l=2$ at smaller internculear separations is also necessary. On the other hand, the centrifugal barrier for $l=3$ is so repulsive that even when the other core is as close as $R=25$\thinspace a.u. from the center of ther perturber, it is not able to reduce the repulsive barrier enough and the peak of $V_{\text{mod}}(z)$ has a positive value. Therefore, it is difficult for an Rydberg electron in the $f$-wave with any negative energy to penetrate into the perturber space defined by $r\lesssim 7$\thinspace a.u.
		
		Of course, the importance of this effect depends on the size of the perturber and on its distance from the positively charged Rydberg core. As can be seen in \figref{fig:penetscheme}, the model potential employed by \textcite{Khuskivadze} for $l=0$ and $l=1$ is suitable and sufficient for the range of the internuclear separations considered in the reference \cite{Khuskivadze}. The contributions with the higher values of $l$ become relevant only at considerably smaller internuclear distances that are the focus of the present study. Two-electron approach to the problem developed in this paper treats the mechanism described above without a need to construct different perturber potentials for different partial waves.
		
		Another important technical difference between the approach developed in \cite{Khuskivadze} and the present paper is the method used to solve the \sche in the inner region. While \textcite{Khuskivadze} perform a direct propagation of the solutions from the center of the sphere to the boundary with careful treatment of the spin-orbit coupling, present work employs the \rmath method, where the logarithmic derivative of the wave function on the sphere is calculated and used to ensure the smooth matching to the solution in the outer region. The spin-orbit effects are not considered in the present work.
		
		This paper is focused on the detailed derivation of the two-electron \rmath approach to the Rydberg molecular states. A simple application to H$_2$ is discussed and the results are compared with previously published simple models to show that it yields correct energies. Its application to Rb$_2$ and other alkali-metal dimers will be subject of a forthcoming paper.
		
		The rest of this paper is organized as follows: The two-electron computational method is derived in Section \ref{sec:rmatmethod}, the numerical aspects of the calculations are discussed in Section \ref{sec:numeri}, the results obtained using this method for H$_2$ are presented in Section \ref{sec:secresults} and Section \ref{sec:coclusions} contains the conclusions. Atomic units are employed throughout the paper, unless stated otherwise.
	\section{Two-electron R-matrix method}
		\label{sec:rmatmethod}
\begin{figure}
			\includegraphics[scale=1.1]{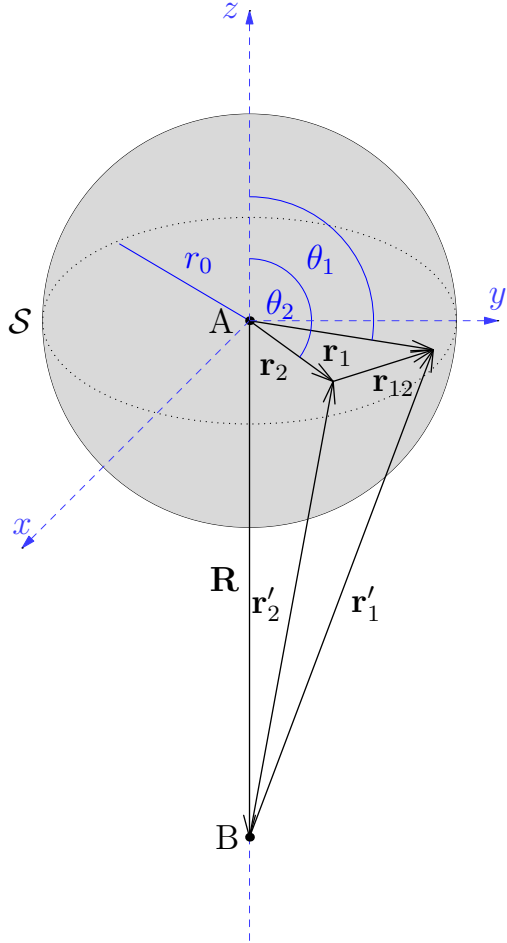}
			\caption{\label{fig:scheme}Coordinate system used in this work with the center on the nucleus of the perturber \cora. The nucleus of the Rydberg core \corb is located at $\mathbf{R}=(0,0,-R)$. Position vectors $\mathbf{r}_1$ and $\mathbf{r}_2$ centered on the origin describe the coordinates of the first and second electron, respectively. Position vectors with respect to the nucleus of the core \corb are denoted as $\mathbf{r}'_1$ and $\mathbf{r}'_2$.}
		\end{figure}

		Following \textcite{Khuskivadze} and \textcite{Hamilton-PhD}, let us introduce a coordinate system with the center located on the nucleus of the neutral perturber \cora (\figref{fig:scheme}). The other positively charged center \corb of the Rydberg atom is located on the negative part of the axis $z$ and its distance from \cora is denoted as $R$. Axis $z$ is also the quantization axis. Position vectors of the electrons are denoted $\mathbf{r}_1$ and $\mathbf{r}_2$.
		The atomic cores \cora and \corb are represented by the spherical model potentials $V_A(r)$ and $V_B(r')$, respectively. The prime in the coordinate denotes that it is taken relatively to the center \corb, while the unprimed coordinate is relative to the core \cora. Both $V_A$ and $V_B$ have attractive Coulomb character beyond certain distance from the corresponding core, typically few atomic units. The short-range part of the potential $V_A$ is optimized 
		to reproduce the energies of the ground and excited states of the perturber with respect to the atomic ionization threshold. Although $V_A(r)$ can be $l$-dependent (see~\cite{Aymar1996} and references therein), such dependence is not needed in this work.
		
		The short-range part of $V_B$ does not explicitly appear in the calculation. Interaction of the Rydberg electron with core $B$ is parametrized by the quantum defects \cite{Davydkin}.
		Electronic Hamiltonian of the whole system is
		\begin{eqnarray}
			\label{eq:totham}
			\hat{H}&=&\hat{K}_1+V_A(r_1)+V_B(\left|\mathbf{r}_1-\mathbf{R}\right|)\nonumber\\*
			       &+&\hat{K}_2+V_A(r_2)+V_B(\left|\mathbf{r}_2-\mathbf{R}\right|)+\frac{1}{r_{12}},
		\end{eqnarray}
		where $\hat{K}_{1}=-(1/2)\nabla_{1}^2$ and $\hat{K}_{2}=-(1/2)\nabla_{2}^2$ are the operators of the kinetic energy for electrons 1 and 2, respectively. The last term is the Coulomb repulsion of the electrons and $r_{12}=\left|\mathbf{r}_2-\mathbf{r}_1\right|$.
		
		This work deals with such electronic states of the two-electron system, where at least one electron always appears close to the core \cora (valence electron) and where the internuclear separation $R$ is large enough so that all the region of the valence electron is located in the Coulomb tail of $V_B$ and does not overlap with the core \corb. The Rydberg electron is asymptotically bound by $V_B$ and its wave function may reach or even exceed the perturber.
		
		Therefore, it is reasonable to introduce the sphere $\mathcal{S}$ centered on the nucleus of the perturber with such radius $r_0$ that whole the region of the space, where the two electrons can appear simultaneously, is confined inside $\mathcal{S}$ (\figref{fig:scheme}).
		The basis for the \rmath formulation of the problem is provided by the separation of the space to two regions. The region inside the sphere $\mathcal{S}$ accounts for an antisymmetry of the two-electron wave function and the effects of the electron correlation. The region outside the sphere $\mathcal{S}$ deals with a single Rydberg electron and matching of its wave function with the two-electron wave function in the inner region can be done in terms of the Coulomb \gfn \cite{Hostler1963}.
		This imposes the proper boundary conditions and treats the Coulomb singularity analytically. Therefore, it is not necessary to construct any basis set to represent the complicated spatial structure of the one-electron wave function outside $\mathcal{S}$ that spreads over relatively large region of the space.
		
		The restriction that both electrons can simultaneously appear only inside $\mathcal{S}$ and at most one of them around the core \corb, leads to an asymmetry in the representation of the studied system. It means that even in the case of the homonuclear molecule the computational method discussed here cannot correctly represent the gerade or ungerade symmetry of the two-electron wave function. That may seem like an important deficiency of this computational approach.  However, one should keep in mind that it is designed for sufficiently large internuclear distances $R$ and for such energy range, where the gerade and ungerade states of the homonuclear systems form nearly degenerate pairs \cite{Greene-prl,Hamilton2002}. The states where both electrons can simultaneously appear only in the vicinity of the atomic core \cora, can be represented by the linear combinations of those degenerate wave functions. The situation is different in case of heteronuclear systems, since it is usually obvious, for the given energy range of the interest, which atom is in the valence state and which is in the Rydberg state.
		
		Let us consider now the low bound states of the diatomic cation system as described above, i.e. in the absence of the Rydberg electron. When $R$ is sufficiently large, the bound state is predominately determined by the potential $V_A(r)$ and the effect of the Coulomb tail of $V_B(\left|\mathbf{r}-\mathbf{R}\right|)$ has only perturbative character. Corresponding one-particle Hamiltonian
		\begin{equation}
			\hat{H}_c=\hat{K}_1+V_A(r_1)+V_B(\left|\mathbf{r}_1-\mathbf{R}\right|)
			\label{eq:chanham}
		\end{equation}
		commutes with the projection of the angular momentum on the internuclear axis $\hat{l}_{1z}$. Therefore, the eigenvalue $m_1$ of $\hat{l}_{1z}$ is a good quantum number and it can be used to categorize the eigenstates of $\hat{H}_c$. Let us assume that $r_0$ is large enough to confine few lowest solutions $\varphi_{im_1}(\mathbf{r}_1)$ of the corresponding Schr\"odinger equation
		\begin{equation}
			\hat{H}_c\varphi_{im_1}(\mathbf{r}_1)=\epsilon_{im_1}\varphi_{im_1}(\mathbf{r}_1)\label{eq:perchanfuncs}
		\end{equation}
		with the bound-state boundary conditions. These eigenstates can represent the scattering channels of the problem. The orthonormality of $\varphi_{im_1}(\mathbf{r}_1)$ is assumed in the rest of this paper. When it is desirable to express the parametric dependence of the channel energies on the internuclear separation $R$, it will be denoted as $\epsilon_{im_1}(R)$ in the remaining part of this paper.
		\subsection{Inner region}
			\label{sec:innerreg}
			In order to calculate the \rmat on the sphere $\mathcal{S}$ variationally, we expand the solutions $\Psi(\mathbf{r}_1,\mathbf{r}_2)$ of the \sche
			\begin{equation}
				\hat{H}\Psi(\mathbf{r}_1,\mathbf{r}_2)=E\Psi(\mathbf{r}_1,\mathbf{r}_2)\label{eq:2esche}
			\end{equation}
			in the inner region in terms of the basis functions.
			In the present study we choose antisymmetric two-electron basis functions coupled to form a state of definite total angular momentum $L$, total electronic spin $S$ and parity $P$. This choice for basis set may seem detrimental since $L$ is not a good quantum number of the diatomic system. However, the only term in the Hamiltonian \eqref{eq:totham} that violates the rotational symmetry is the potential $V_B$ that is a smooth function inside the sphere $\mathcal{S}$. We will show that $V_B$ generates in the inner region a weak $L$-coupling that can be expressed by a finite series expansion of the multipole type. One-particle version of this approach has been utilized by \textcite{Khuskivadze}, where the off-center Coulomb potential couples the eigenstates of the operator $(\hat{L}+\hat{S})^2$, since the $jj$-coupling scheme has been employed by the authors. The theoretical studies of the photoionization spectroscopy of two-electron Rydberg atoms (see~\cite{Aymar1996} and references therein) are based on similar approach, where the dipole electric field couples different eigenstates of $\hat{L}^2$.
			
			Following \textcite{Aymar1996}, the two-electron basis function is expressed in terms of the one-electron radial orbitals and angular spherical harmonics:
			\begin{eqnarray}
				y_{n_1l_1n_2l_2}^{(LM)}(\mathbf{r}_1,\mathbf{r}_2)&=&C_{n_1l_1n_2l_2}\left[\vphantom{\mathcal{Y}_{l_2l_1}^{(LM)}}u_{n_1l_1}(r_1)f_{n_2l_2}(r_2)\right.\nonumber\\*
				\times\mathcal{Y}_{l_1l_2}^{(LM)}(\mathbf{\Omega}_1,\mathbf{\Omega}_2)&+&(-1)^{l_1+l_2+L-S}f_{n_2l_2}(r_1)u_{n_1l_1}(r_2)\nonumber\\*
				&\times&\left.\mathcal{Y}_{l_2l_1}^{(LM)}(\mathbf{\Omega}_1,\mathbf{\Omega}_2)\right]/(r_1r_2),\label{eq:2ebasis}
			\end{eqnarray}
			where $l_1$ and $l_2$ are the angular momenta of the first and second electron, respectively. Eigenvalue $M$ corresponds to the projection of the total angular momentum on the quantization axis $\hat{L}_z$. The indices $n_1$ and $n_2$ represent the one-particle radial orbitals. The set of indices $\{n_1,l_1,n_2,l_2\}$ will be denoted by a joint index $\gamma$ in the rest of this paper. $\mathbf{\Omega}_1=(\theta_1, \phi_1)$ and $\mathbf{\Omega}_2=(\theta_2, \phi_2)$ are the angular coordinates of the first and the second electron, respectively (see \figref{fig:scheme}). The two-electron angular functions $\mathcal{Y}_{l_1l_2}^{(LM)}(\mathbf{\Omega}_1,\mathbf{\Omega}_2)$ are the spherical harmonics in the angular coordinates of both electrons coupled by the Clebsch-Gordon coefficients to form the eigenstates of $\hat{L}^2$ and $\hat{L}_z$, as it is usual in the addition of two angular momenta~\cite{edmonds,Sobelman1972}. $C_{n_1l_1n_2l_2}=1/\sqrt{2(1+\delta_{n_1n_2}\delta_{l_1l_2})}$ is the normalization factor~\cite{Aymar1996}. Since $M$ is the conserved quantum number of the system studied here, it will be skipped from the notation of the basis functions \eqref{eq:2ebasis} and corresponding matrix elements in the rest of the paper. The desired value of $M$ is chosen at the beginning of the calculation as the parameter and it has the same fixed value in all the equations in this paper.
			
			The radial orbitals $u_{nl}(r)$, $n=1\dots N_c$ represent the bound states of the bare neutral perturber, $N_c$ is the number of functions employed in the basis set \eqref{eq:2ebasis} for each value of the angular momentum $l$. One-electron orbitals $u_{nl}(r)$ are calculated variationally as linear combinations of the elements of the \bspl basis set~\cite{Bachau2001}
			\begin{equation}
				\label{eq:bexcoefs}
				u_{nl}(r)=\sum_{\alpha=1}^{N_b}b_{\alpha nl}B_\alpha(r),
			\end{equation}
			where $B_\alpha(r)$ are the \bspl basis functions defined on the interval $\left[0,r_0\right]$ and $b_{\alpha nl}$ are the expansion coefficients. Substitution of \eqab{eq:bexcoefs} into the radial \sche
			\begin{subequations}
				\label{eq:batorbs}
				\begin{eqnarray}
					\left[-\frac{1}{2}\frac{d^2}{dr^2}+\frac{l(l+1)}{2r^2}+V_A(r)\right]u_{nl}(r)&=&\varepsilon_{nl}u_{nl}(r)\label{eq:batsche}\\*
					u_{nl}(0)&=&0\label{eq:bartorbsbcin}\\*
					u_{nl}(r_0)&=&0\label{eq:bartorbsbcout},
				\end{eqnarray}
			\end{subequations}
			its projection onto the basis functions and solution of the obtained generalized eigenvalue problem for each $l$ (since the \bspls form a non-orthogonal basis set) then yields the eigenvalues $\varepsilon_{nl}$ and coefficients $b_{\alpha nl}$. The eigenvectors are normalized so that
			\begin{equation}
				\int\limits_{0}^{r_0}dr\,u_{nl}(r)u_{n'l}(r)=\delta_{nn'}.
			\end{equation}
			The radius $r_0$ of $\mathcal{S}$ is chosen large enough so that the effect of the finite box can be neglected for lowest states. This is the usual requirement of the \rmath methods.
			
			The set of the radial orbitals $f_{n_2l_2}(r)$ that appear in \eqab{eq:2ebasis} consists of the orbitals $u_{nl_2}(r)$ and it is completed by the set of \bspls as
			\begin{equation}
				f_{n_2l_2}(r)=
				\begin{dcases}
					u_{nl_2}(r) & n=1\dots N_c,\;l_2=0,1,\dots\\
					B_{\alpha}(r) & \alpha=1\dots N_b
				\end{dcases}
				.
			\end{equation}
			Since all the orbitals $u_{nl}(r)$ vanish on the sphere $\mathcal{S}$, they are denoted as ``closed'' in the following text. On the other hand, the last \bspl $B_{N_b}(r)$ has non-zero value at $r=r_0$. Hence, this subset represents the ``open'' orbitals that are neither orthogonal to one another nor to $u_{nl}(r)$. The two-particle basis function \eqref{eq:2ebasis}, where $f_{nl}$ is $u_{nl}$ ($B_n$), is denoted closed (open). The absence of the two-particle basis functions with two electrons in the open orbitals reflects that at most one electron can escape from the reaction volume.
			
			While the open and closed categories of the radial orbitals and two-particle basis functions introduced above have the same meaning as in the reference \cite{Aymar1996}, the character of the open subset is very different. \textcite{Aymar1996} obtained all the radial orbitals by the numerical integration of \eqab{eq:batsche} on the grid. The closed orbitals $u_{nl}(r)$ were calculated in such way that the boundary conditions \eqref{eq:bartorbsbcin} and \eqref{eq:bartorbsbcout} were imposed. These closed orbitals are similar to those introduced in this work, only the method of integration is different. However, the open orbitals in the reference \cite{Aymar1996} were obtained by the numerical integration of the differential equation \eqref{eq:batsche} with the boundary condition in the center \eqref{eq:bartorbsbcin} for arbitrary energy $\varepsilon_{nl}$ without imposing any boundary condition at $r_0$. Of course, the selection of the energy was physically motivated. In contrast to the work by \textcite{Aymar1996}, the open orbitals introduced in this paper do not obey any differential equation and simply form a complete set of functions on the interval $\left[0,r_0\right]$. As a result, the construction of the open orbitals is less ambiguous than in the reference \cite{Aymar1996}, since it does not involve any other parameter. The numerical aspects of this technique are discussed below. Note that similar \rmath method to treat two-electron atoms based on the \bspl functions has been developed by \textcite{Hart1997}. \textcite{Zatsarinny} has employed \bspls as a continuum basis set in his \rmath calculations of the electron-atom collisions as well. That method and corresponding computer program is capable of treatment of much more complex atoms with many electrons.
			
\begin{figure}
				\includegraphics[scale=0.87]{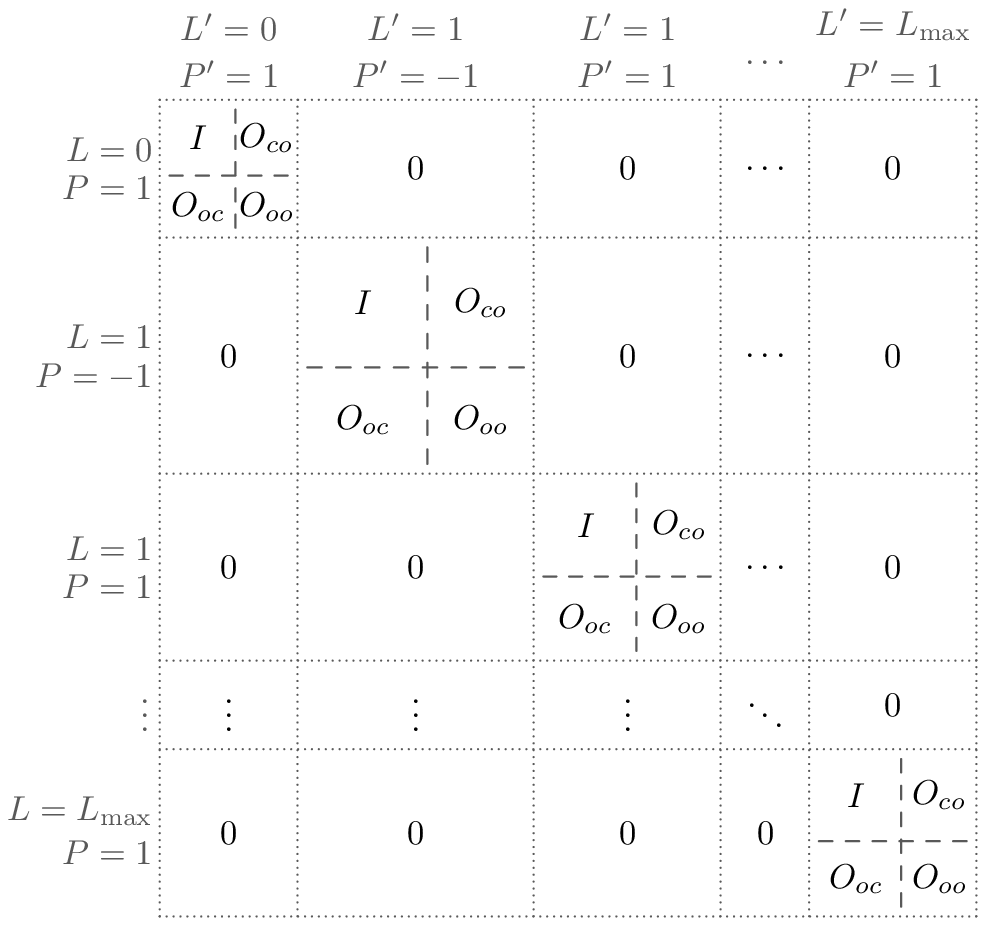}
				\caption{Block structure of the overlap matrix $\underline{O}$ (see \eqab{eq:2eovermatels}). All the blocks where $L\neq L'$ are zero. The mutual overlaps of the closed basis functions yield the identity matrices $O_{cc}=I$ in the diagonal blocks.\label{fig:omatscheme}}
			\end{figure}

\begin{figure}
				\includegraphics[scale=0.87]{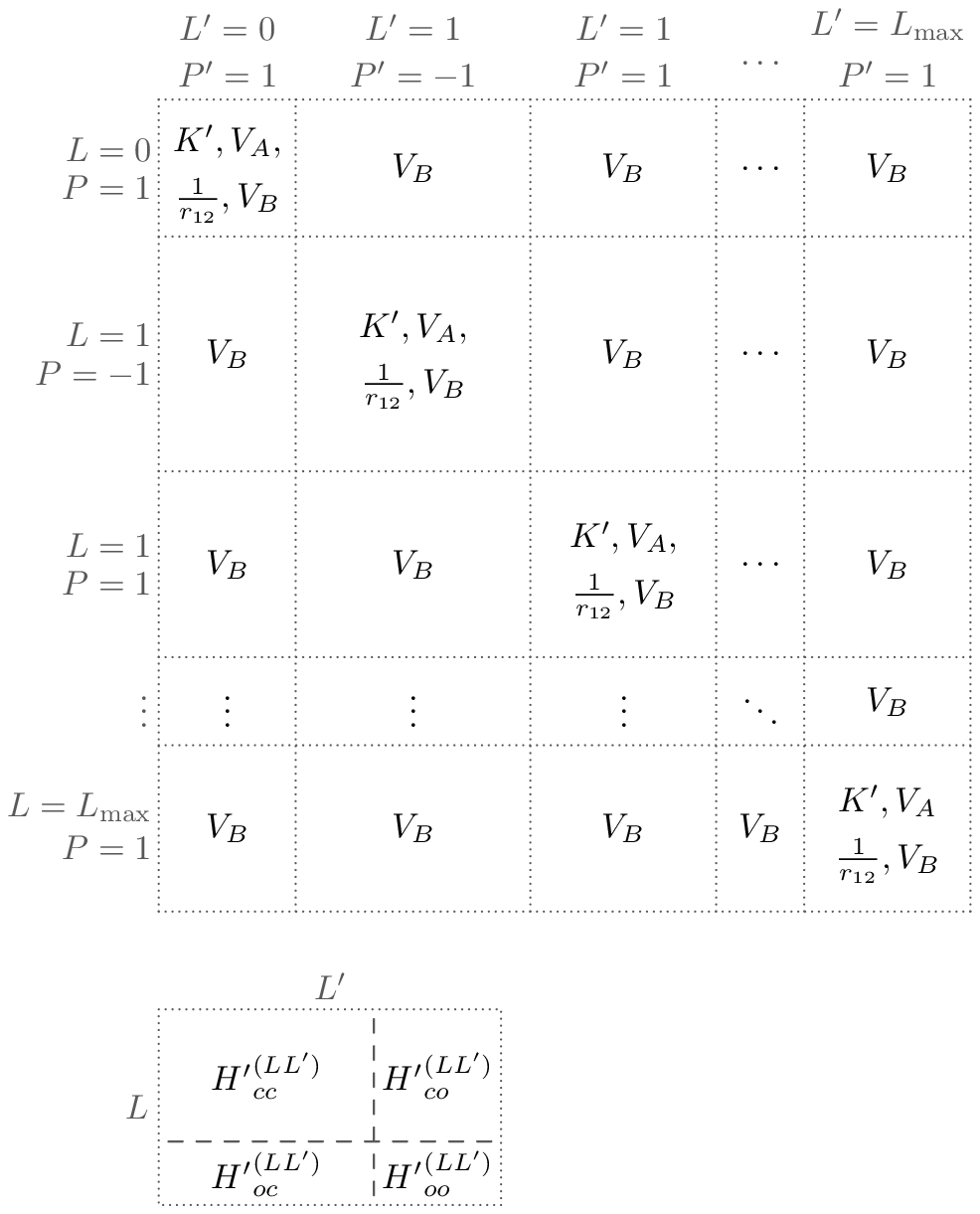}
				\caption{Block structure of the Hamiltonian matrix $\underline{H}'$. All the non-zero matrix elements of the operators $\hat{K}'=\hat{K}_1+\hat{K}_2+\hat{L}_{B}$, $\hat{V}_A=V_A(r_1)+V_A(r_2)$ and $1/r_{12}$ are located only in the diagonal blocks. The non-zero matrix elements of $\hat{V}_B=V_B(r_1')+V_B(r_2')$ are not restricted to any specific blocks. The small scheme on the bottom of the figure shows that every $LL'$-block consists of the sub-matrices corresponding to groupping the basis functions into the open and closed set.\label{fig:hamscheme}}
			\end{figure}

			Using the basis set \eqref{eq:2ebasis}, the variational principle for the \rmat can be formulated. We employ the Wigner-Eisenbud pole expansion of the \rmat \cite{Robicheaux1991, Aymar1996}, where the Bloch operator \cite{Bloch1957}
			\begin{equation}
				\hat{L}_B=\frac{1}{2r_1}\delta(r_1-r_0)\frac{d}{dr_1}r_1+\frac{1}{2r_2}\delta(r_2-r_0)\frac{d}{dr_2}r_2
			\end{equation}
			is added to the Hamiltonian \eqref{eq:totham} and the matrix representation $\underline{H}'$ of this modified operator $\hat{H}'=\hat{H}+\hat{L}_B$ in the basis set \eqref{eq:2ebasis} is diagonalized. The matrix elements
			\begin{eqnarray}
				O_{\gamma\gamma'}^{(LL')}&=&\Braket{y_\gamma^{(L)}|y_{\gamma'}^{(L')}}\label{eq:2eovermatels}\\
				{H'}_{\gamma\gamma'}^{(LL')}&=&\Braket{y_\gamma^{(L)}|\hat{H}'|y_{\gamma'}^{(L')}}\label{eq:2ehammatels},
			\end{eqnarray}
			involve an integration throughout the inner region in both coordinates $\mathbf{r}_1$ and $\mathbf{r}_2$. Both matrices $\underline{H}'$ and $\underline{O}$ are in our calculations organized into blocks defined by corresponding $L$ and $L'$. Each block has a structure corresponding to groupping the basis functions \eqref{eq:2ebasis} into the open and closed subset (see \figref{fig:omatscheme} and \figref{fig:hamscheme}). Note that the non-orthogonality of the open orbitals $f_{nl}(r)$ is transfered to the two-electron basis functions \eqref{eq:2ebasis}. Therefore, the overlap matrix $\underline{O}$ defined by \eqab{eq:2eovermatels} is block diagonal with $L=L'$. The closed-closed parts of the non-zero blocks are formed by the identity matrices $I$ (\figref{fig:omatscheme}).
						
			Evaluation of the matrix elements \eqref{eq:2ehammatels} for the one-particle operators $\hat{K}_{1,2}$, $V_A$ and $\hat{L}_B$ is straightforward. We first calculate their representation in the \bspl basis  and then perform the contraction \eqab{eq:bexcoefs}. The matrix elements in the two-electron basis \eqref{eq:2ebasis} can be expressed in terms of these one-electron elements, since the angular integration is trivial and can be performed analytically. Note that all these operators are diagonal in $L$. Therefore, when matrix $\underline{H}'$ is organized into blocks indexed in rows and columns by $L$ and $L'$, respectively, all the non-zero matrix elements of operators $\hat{K}_{1,2}$, $V_A$ and $\hat{L}_B$ are located only in the diagonal blocks $L=L'$ (see \figref{fig:hamscheme}).
			
			The operator of the electron repulsion $1/r_{12}$ is also diagonal in $L$. However, calculation of its matrix elements is computationally more demanding than evaluation of the one-particle terms discussed above. As it is usual in the theory of atomic spectra (for example, see \cite{Sobelman1972,Zatsarinny,Aymar1996} and references therein), the electron repulsion operator is first expressed in terms of the multipole series
			\begin{equation}
				\frac{1}{\left|\mathbf{r}_1-\mathbf{r}_2\right|}=\sum\limits_{\lambda=0}^{\infty}\frac{4\pi}{2\lambda+1}\frac{r_{<}^\lambda}{r_{>}^{\lambda+1}}\sum\limits_{\mu=-\lambda}^{\lambda}Y^*_{\lambda\mu}(\mathbf{\Omega}_1)Y_{\lambda\mu}(\mathbf{\Omega}_2),
				\label{eq:polexpand}
			\end{equation}
			where $r_<$ ($r_>$) is the smaller (larger) value of $r_1$ and $r_2$, while $Y_{\lambda\mu}(\mathbf{\Omega})$ are the spherical harmonics. The multipole expansion \eqref{eq:polexpand} allows to carry out the angular integration analytically \cite{Sobelman1972}. The radial terms are calculated using the elements
			\begin{eqnarray}
				R^\lambda(\alpha_1,\alpha_2,\alpha_1',\alpha_2')&=&\int\limits_{0}^{r_0}dr_1\int\limits_{0}^{r_0}dr_2\,B_{\alpha_1}(r_1)B_{\alpha_1'}(r_1)\nonumber\\*
				&\times&\frac{r_<^\lambda}{r_>^{\lambda+1}}B_{\alpha_2}(r_2)B_{\alpha_2'}(r_2)
			\end{eqnarray}
			that are numerically integrated by the cell method proposed by \textcite{Qiu}. These integrals are then contracted with the coefficients $b_{\alpha nl}$ in two, three or four indices, depending on whether the matrix element couples the open-open, open-closed or closed-closed pairs of basis functions \eqref{eq:2ebasis}, respectively. It is this contraction that makes the evaluation of the matrix elements computationally demanding, since in case of the transformation of all four indices the number of necessary multiplications and additions scales with the dimension of the \bspl basis as $N_b^5$ \cite{Diercksen}. Even though the number of orbitals $u_{nl}(r)$ used in construction of the basis functions \eqref{eq:2ebasis} is in practical calculations smaller than the dimension of the \bspl basis ($N_c<N_b$), this step is a bottleneck in the numerical construction of $\underline{H}'$. The same type of transformation is also employed in the \textit{ab-initio} programs for the molecular electronic structure calculations, when the four-index integrals calculated in the atomic basis are transformed to the molecular basis \cite{Diercksen}.
			
			Note that thanks to the independence of the two-electron basis functions \eqref{eq:2ebasis} on $R$, the matrix elements of all the operators discussed above do not depend on the internuclear separation $R$. Therefore, they are calculated only once and used in calculations of the energy spectra at each internuclear separation of the interest.
			
			The only part of the Hamiltonian $\hat{H}$ that depends on $R$, is the potential $V_B(r')$ of the Rydberg atomic core $B$. However, the evaluation of corresponding matrix elements in basis set \eqref{eq:2ebasis} is computationally tractable compared to the matrix elements of the electron-electron repulsion. Assuming that whole the atomic core \corb is in the outer region, the potential $V_B(r')$ inside the sphere $\mathcal{S}$ has form $V_B(\left|\mathbf{r}-\mathbf{R}\right|)=-1/\left|\mathbf{r}-\mathbf{R}\right|$ and its corresponding matrix elements can be calculated using the pole expansion \eqref{eq:polexpand} as it is explained in Appendix \ref{sec:append}.
			
			Having the Hamiltonian matrix (modified by the Bloch term) $\underline{H}'$ and the overlap matrix $\underline{O}$, the \rmat can be expressed in terms of the solutions of the generalized eigenvalue problem \cite{Aymar1996, Robicheaux1991, Zatsarinny}
			\begin{equation}
				\underline{H}'\mathbf{c}_k=E_k\underline{O}\mathbf{c}_k,
				\label{eq:geneigen}
			\end{equation}
			where $E_k$ are the eigenvalues and $\mathbf{c}_k$ are corresponding (column) eigenvectors normalized by the overlap matrix to \cite{Robicheaux1991}
			\begin{equation}
				\mathbf{c}_k^T\underline{O}\mathbf{c}_{k'}=\delta_{kk'}.
			\end{equation}
			In order to construct the \rmat, let us first express the general solutions of the \sche \eqref{eq:2esche} in the outer region as a channel expansion
			\begin{equation}
				\left.\Psi_\beta(\mathbf{r}_1,\mathbf{r}_2)\right|_{r_2\geq r_0}=\sum\limits_{im_1}\varphi_{im_1}(\mathbf{r}_1)Q_{im_1\beta}(\mathbf{r}_2),\label{eq:chanexpand}
			\end{equation}
			where $\beta$ denotes different independent solutions at the same energy $E$. These solutions usually obey some specific boundary conditions in the outer region (i.e. bound state, linear combination of the incoming and outgoing wave, etc.). When we denote $m_2\equiv M-m_1$ and the scattering wave function $Q_{im_1\beta}(\mathbf{r}_2)$ in the outer region is expressed as 
			\begin{equation}
				\left.Q_{im_1\beta}(\mathbf{r}_2)\right|_{r_2\geq r_0}=\sum\limits_{l_2=\left|m_2\right|}^\infty\frac{1}{r_2}q_{\bar{j}\beta}(r_2)Y_{l_2m_2}(\mathbf{\Omega}_2),
				\label{eq:fexpand}
			\end{equation}
			where $\bar{j}=\{i,m_1,l_2\}$ is the scattering channel, then the \rmat couples $q_{\bar{j}\beta}(r_0)$ evaluated on the sphere $\mathcal{S}$ with their radial derivatives
			\begin{equation}
				q_{\bar{j}\beta}(r_0)=\sum\limits_{\bar{j}'}R_{\bar{j}\bar{j}'}q_{\bar{j}'\beta}'(r_0).
				\label{eq:rmatcond}
			\end{equation}
			The notation $q_{\bar{j}'\beta}'(r_0)$ for the derivative of the radial wave function $q_{\bar{j}'\beta}(r)$ evaluated on the sphere $\mathcal{S}$ introduced in \eqab{eq:rmatcond} will be utilized in the remaining part of this paper also for other wave functions.
			The elements of the \rmat can be expressed as \cite{Aymar1996,Robicheaux1991}
			\begin{equation}
				R_{\bar{j}\bar{j}'}(E)=\frac{1}{2}\sum\limits_{k}\frac{w_{\bar{j}k}w_{\bar{j}'k}}{E_k-E}.
				\label{eq:rmatpolex}
			\end{equation}
			The surface amplitudes $w_{\bar{j}k}$ can be calculated using the eigenvectors $\mathbf{c}_k$. Their explicit form is derived in Appendix \ref{sec:amplider}. The channel index $\bar{j}$ characterizes the valence state of the perturber as well as the partial wave of the electron that can escape from the inner region. Universal validity of the boundary condition \eqref{eq:rmatcond} is the strong side of the \rmath method, as the set of solutions $q_{\bar{j}\beta}(r_0)$ in \eqab{eq:rmatcond} is not restricted by any other boundary condition on $\mathcal{S}$ nor in the outer region.
			
			The solution of the generalized eigenvalue problem \eqref{eq:geneigen} is numerically most demanding step of all the calculation. It is necessary to perform a complete diagonalization of the modified Hamiltonian $\underline{H}'$ as all the eigenvalues and eigenvectors are required to construct the \rmat \eqref{eq:rmatpolex}. Apparently, it is necessary to perform this step for each internuclear distance $R$ of the interest. Note, however, that once having the eigenrepresentation of $\underline{H}'$, the dependence of the pole expansion of the \rmat on the energy \eqref{eq:rmatpolex} is very simple. This is also the main difference between the present method and another very powerful technique - the eigenchannel \rmat developed by \textcite{Greene1988}, where the \rmat can be evaluated by solving a system of linear algebraic equations \cite{Aymar1996}, however, the set of linear equations must be solved for each energy of the interest. Therefore, the method employed here is more convenient, since the search for the bound states is performed on relatively fine grid of energies.
			
		\subsection{Outer region}
			\label{sec:outerreg}
			One-electron equations (for the coordinate $\mathbf{r}_2$) in the outer region can be obtained by substitution of the solution \eqref{eq:chanexpand} into the \sche \eqref{eq:2esche}. Resulting equation is then multiplied by the wave functions $\varphi_{im_1}^*(\mathbf{r}_1)$ and integrated over the coordinate $\mathbf{r}_1$ throughout the inner region. We arrive at the coupled set of differential equations
			\begin{eqnarray}
				\left[\hat{K}_2\right.&+&\left.\vphantom{\hat{K}_2+}V_B(\left|\mathbf{r}_2-\mathbf{R}\right|)-(E-\epsilon_{im_1})\right]Q_{im_1\beta}(\mathbf{r}_2)\nonumber\\*
				&=&\sum_{i'm_1'}U_{im_1i'm_1'}(\mathbf{r}_2)Q_{i'm_1'\beta}(\mathbf{r}_2),\label{eq:outersche}
			\end{eqnarray}
			where the coupling potential has the following form:
			\begin{eqnarray}
				U_{im_1i'm_1'}(\mathbf{r}_2)=&-&\int_{\mathcal{V}}dV_1\,\varphi_{im_1}^*(\mathbf{r}_1)\frac{1}{r_{12}}\varphi_{i'm_1'}(\mathbf{r}_1)\nonumber\\*
				&+&\frac{1}{r_2}\delta_{ii'}\delta_{m_1m_1'}.\label{eq:outerpoten}
			\end{eqnarray}
			The domain of integration $\mathcal{V}$ in \eqab{eq:outerpoten} is the volume enclosed by $\mathcal{S}$. The second term reflects the fact the $V_A(r)=-1/r$ for $r>r_0$.
			
			The first term in \eqab{eq:outerpoten} represents the repulsion of the Rydberg electron with the valence electron confined inside the sphere $\mathcal{S}$. The second term describes the attraction between the electron in the outer region with the core \cora. When we substitute the multipole expansion \eqref{eq:polexpand} into the first term, we immediately see that the first (charge) term of the expansion cancels with the second term of \eqab{eq:outerpoten}. The first non-vanishing term of \eqab{eq:outerpoten} is the dipole potential. It does not vanish because the wave functions $\varphi_{im_1}(\mathbf{r}_1)$ are not typical atomic states. They are eigenstates of the atomic Hamiltonian perturbed by the Coulomb tail of the potential $V_B$ \eqref{eq:chanham}. Hence the non-vanishing dipole term is a consequence of the polarization of the valence electron by the core \corb. Explicit form of this induced dipole moment is given in Appendix \ref{sec:dipappend}. The induced dipole field was previously employed to model the polarization of the neutral perturber \cite{Khuskivadze} or its core \cite{bottcher, Henriet1984} by other positively charged atomic center. In Section \ref{sec:numeri} we will analyze its magnitude and its dependence on the internuclear distance $R$. Our goal is to provide a numerical evidence that will allow us to neglect the coupling potential on the right-hand side of \eqab{eq:outersche} outside the sphere $\mathcal{S}$.

			Provided that the right-hand side of \eqab{eq:outersche} can be neglected, the remaining system of differential equations is not coupled in the indices $i$, $m_1$ and $\beta$. The bound-state energies of the studied system can be found by solving the homogeneous part of \eqab{eq:outersche} for $r_2>r_0$ with the boundary condition \eqref{eq:rmatcond} on the sphere $\mathcal{S}$. Only those solutions that vanish asymptotically are chosen. The two boundary conditions can be satisfied simultaneously only for a discrete set of energies. Since the coordinates of the valence electron do not appear in the rest of this section, the index of the electron position vector in the outer region is dropped and the coordinates are denoted as $\mathbf{r}=(r,\theta,\phi)$. Any solution $X_{im_1}(\mathbf{r})$ that obeys all the desired bound-state boundary conditions can be expressed as a linear combination of the complete set of solutions $Q_{im_1\beta}(\mathbf{r})$ of \eqab{eq:fexpand} as
			\begin{equation}
				X_{im_1}(\mathbf{r})=\sum_{\beta}A_\beta Q_{im_1\beta}(\mathbf{r})\label{eq:bsradwfacomb}
			\end{equation}
			and the relation of the radial components $x_{\bar{j}}(r)$ defined by the equation
			\begin{equation}
				X_{im_1}(\mathbf{r})=\sum_{l_2=\left|m_2\right|}^\infty\frac{x_{\bar{j}}(r)}{r}Y_{l_2m_2}(\mathbf{\Omega})
				\label{eq:outerboundradex}
			\end{equation}
			to the general radial solutions \eqref{eq:fexpand} is
			\begin{equation}
				x_{\bar{j}}(r)=\sum_{\beta}A_\beta q_{\bar{j}\beta}(r).
				\label{eq:aradex}
			\end{equation}
			The bound-state solutions $X_{im_1}(\mathbf{r})$ satisfy the homogeneous part of \eqab{eq:outersche}, i.e.
			\begin{equation}
				\left[-\frac{1}{2}\nabla^2+V_B(\left|\mathbf{r}-\mathbf{R}\right|)-(E-\epsilon_{im_1})\right]X_{im_1}(\mathbf{r})=0.
				\label{eq:nextoutersche}
			\end{equation}
			Methods based on the partial-wave expansion of the solution with respect to the center of the perturber \cora, would be, due to the singularity of $V_B$ at the position of the Rydberg core \corb, unpractical and computationally very demanding. Instead, we employ a straightforward multichannel generalization of the treatment developed by \textcite{Khuskivadze} that is based on the \gfn, originally formulated by \textcite{Fabrikant1993}. Consider the \gfn $G(\mathbf{r},\mathbf{r}'')$ for the potential $V_B$ defined by the equation
			\begin{eqnarray}
				\left[-\frac{1}{2}\nabla^2\vphantom{V_B(\left|\mathbf{r}-\mathbf{R}\right|)-\frac{k_{im_1}^2}{2}}\right.&+&\left.\vphantom{-\frac{1}{2}\nabla^2}V_B(\left|\mathbf{r}-\mathbf{R}\right|)-\frac{k_{im_1}^2}{2}\right]\nonumber\\*
				&\times&G_{im_1}(\mathbf{r},\mathbf{r}'')=-\delta^3(\mathbf{r}-\mathbf{r}''),\label{eq:outergfe}
			\end{eqnarray}
			where $k_{im_1}=\sqrt{2(E-\epsilon_{im_1})}$. We are interested in such $G_{im_1}(\mathbf{r},\mathbf{r}'')$ that satisfies the asymptotic bound-state boundary conditions. If the Rydberg core \corb is the hydrogen core, then $G_{im_1}(\mathbf{r},\mathbf{r}'')$ is the Coulomb \gfn derived in the closed form by \textcite{Hostler1963}. For more complex positive ions, a correction of the Coulomb \gfn developed by \textcite{Davydkin} is added \cite{Khuskivadze} to account for the non-hydrogen character of the core. That correction is then parametrized by the corresponding atomic quantum defects. When we multiply \eqab{eq:outersche} by $G_{im_1}(\mathbf{r},\mathbf{r}'')$, subtract it from \eqab{eq:outergfe} multiplied by $X_{im_1}(\mathbf{r})$ and then integrate it over $\mathbf{r}$, we arrive at
			\begin{eqnarray}
				&-&\frac{1}{2}\int_{\mathcal{S}_0}d\Omega\left[G_{im_1}(\mathbf{r},\mathbf{r}'')\nabla^2X_{im_1}(\mathbf{r})-X_{im_1}(\mathbf{r})\right.\nonumber\\*
				&\times&\left.\nabla^2G_{im_1}(\mathbf{r},\mathbf{r}'')\right]=\int_{\mathcal{S}_0}d\Omega\,\delta^3(\mathbf{r}-\mathbf{r}'')X_{im_1}(\mathbf{r}),
				\label{eq:outvolint}
			\end{eqnarray}
			where the domain of the integration $\mathcal{S}_0$ is the unit sphere. The \gfn can be expanded in terms of the spherical harmonics in the angular coordinates as
			\begin{eqnarray}
				G_{im_1}(\mathbf{r},\mathbf{r}'')&=&\sum_{\lambda=0}^\infty\sum_{\mu=-\lambda}^{\lambda}\sum_{\lambda'=\left|\mu\right|}^\infty\frac{g_{im_1\lambda\lambda'\mu}(r,r'')}{rr''}\nonumber\\*
				&\times&Y_{\lambda\mu}^*(\mathbf{\Omega})Y_{\lambda'\mu}(\mathbf{\Omega}''),
				\label{eq:gfsphex}
			\end{eqnarray}
			where $\mathbf{\Omega}''=(\theta'',\phi'')$ are the angular coordinates of $\mathbf{r}''$ and the symmetry with respect to the rotations around the $z$-axis has been utilized \cite{Khuskivadze}. Substitution of this expansion along with \eqab{eq:outerboundradex} into \eqab{eq:outvolint} and integration over $\mathbf{\Omega}$ yields
			\begin{eqnarray}
				&-&\frac{1}{2}\sum_{\substack{\lambda'\\l_2\in\bar{j}}}\left[g_{\bar{j}\lambda'}(r,r'')\frac{d^2}{dr^2}x_{\bar{j}}(r)-x_{\bar{j}}(r)\frac{\partial^2}{\partial r^2}g_{\bar{j}\lambda'}(r,r'')\right]\nonumber\\*
				&\times&Y_{\lambda'm_2}(\mathbf{\Omega}'')=\sum_{l_2}Y_{l_2m_2}(\mathbf{\Omega}'')x_{\bar{j}}(r)\delta(r-r'').\label{eq:angulint}
			\end{eqnarray}
			The only terms of expansion \eqref{eq:gfsphex} that remain in \eqab{eq:angulint} after the angular integration, are those where $\mu=m_2=M-m_1$ and $\lambda=l_2$. In order to keep the notation simple, index $\mu$ in $g_{im_1\lambda\lambda'\mu}(r,r'')$ was dropped and leading three indices were again merged into the multi-index $\bar{j}=\left\{i,m_1,l_2\right\}$ denoting the scattering channel.
			
			In the outer region ($r>r_0$) the right-hand side of \eqab{eq:angulint} vanishes in the limit $r''\to r_0-$. Both $x_{\bar{j}}(r)$ and $g_{im_1\lambda\lambda'\mu}(r,r'')$ vanish along with their first derivatives for $r\to\infty$ \cite{Hostler1963}. Moreover, the radial integration of \eqab{eq:angulint} can be performed. Subsequent projection on $Y_{\lambda m_2}(\mathbf{\Omega}'')$ then yields
			\begin{equation}
				\sum\limits_{l_2\in\bar{j}}\left[g_{\bar{j}\lambda}(r_0,r_0)x_{\bar{j}}'(r_0)-x_{\bar{j}}(r_0)g_{\bar{j}\lambda}'(r,r_0)\right]=0,
			\end{equation}
			where
			\begin{equation}
				g_{\bar{j}\lambda}'(r_0,r_0)=\lim\limits_{r''\to r_0-}\left.\frac{\partial}{\partial r}\right|_{r=r_0}g_{\bar{j}\lambda}(r,r'').
			\end{equation}
			Boundary condition \eqref{eq:rmatcond} can be employed at this point to establish the relation between the solutions on the sphere and their derivative in order to smoothly match the solutions in the inner region with the solutions outside $\mathcal{S}$:
			\begin{equation}
				\sum\limits_{\substack{\bar{j}'\\l_2\in\bar{j}}}\left[g_{\bar{j}\lambda}(r_0,r_0)\delta_{\bar{j}\bar{j}'}-R_{\bar{j}\bar{j}'}g_{\bar{j}\lambda}'(r_0,r_0)\right]
				x_{\bar{j}'}'(r_0)=0.\label{eq:bselcond}
			\end{equation}
			\eqab{eq:bselcond} represents a necessary condition for the set of the radial coefficients $x_{\bar{j}}(r_0)$ to form a bound-state wave function via \eqab{eq:bsradwfacomb} and to possess a fixed logarithmic derivative (or \rmat $R_{\bar{j}\bar{j'}}$) on the sphere $\mathcal{S}$. The coefficients $x_{\bar{j}}(r_0)$ are in turn expressed as linear combinations of the general solutions $q_{\bar{j}\beta}(r_0)$ (see \eqab{eq:aradex}) that satisfy the boundary condition \eqref{eq:rmatcond} and possess boundary condition of any choice in the outer region. It is convenient to fix the boundary conditions by choosing the derivatives on the sphere $\mathcal{S}$
			\begin{equation}
				q_{\bar{j}\beta}'(r_0)=\delta_{\bar{j}\beta}\label{eq:outbasdiag}.
			\end{equation}
			Substitution of \eqab{eq:aradex} and \eqab{eq:outbasdiag} into \eqab{eq:bselcond} yields the system of linear equations for coefficients $A_\beta$
			\begin{equation}
				\sum\limits_\beta M_{\bar{j}\beta}A_\beta=0,
			\end{equation}
			where the matrix elements of $\underline{M}$ are the expressions in the square brackets in \eqab{eq:bselcond}. More specifically, introducing the matrices $\underline{\Gamma}$ and $\underline{\Gamma}'$ as
			\begin{subequations}
				\label{eq:gmatrices}
				\begin{eqnarray}
					\Gamma_{\bar{j}\bar{j'}}&=&\int_{\mathcal{S}}r_0^3\,d\Omega\int_{\mathcal{S}}r_0^3\,d\Omega''\,G_{im_1}(\mathbf{r},\mathbf{r}'')\nonumber\\*
					&\times&Y_{l_2'm_2}^*(\mathbf{\Omega}'')Y_{l_2m_2}(\mathbf{\Omega})\delta_{ii'}\delta_{m_1m_1'}=\nonumber\\*
					&=&\delta_{ii'}\delta_{m_1m_1'}g_{\bar{j}l_2'}(r_0,r_0)\\
					\Gamma_{\bar{j}\bar{j'}}'&=&\int_{\mathcal{S}}r_0^2\,d\Omega\int_{\mathcal{S}}r_0^3\,d\Omega''\,\lim\limits_{r''\to r_0-}\left.\frac{\partial}{\partial 	r}\right|_{r_0}\left[rG_{im_1}(\mathbf{r},\mathbf{r}'')\right]\nonumber\\*
					&\times&Y_{l_2'm_2}^*(\mathbf{\Omega}'')Y_{l_2m_2}(\mathbf{\Omega})\delta_{ii'}\delta_{m_1m_1'}=\nonumber\\*
					&=&\delta_{ii'}\delta_{m_1m_1'}g_{\bar{j}l_2'}'(r_0,r_0),
				\end{eqnarray}
			\end{subequations}
			matrix $\underline{M}$ can be written as
			\begin{equation}
				\underline{M}=\underline{\Gamma}-\underline{\Gamma'}\,\underline{R}
			\end{equation}
			and the energies of the bound states can be found by solving the equation $\det(\underline{M})=0$.
			
			 Computationally most demanding step in the treatment of the outer region is the construction of matrices $\underline{\Gamma}$ and $\underline{\Gamma}'$. Although the angular integrals in \eqab{eq:gmatrices} can be reduced to three dimensions \cite{Khuskivadze,Hostler1963}, evaluation of the \gfn along with its normal derivative is the bottleneck of the outer region calculation. It is necessary to evaluate it on the grid of the energies $(E-\epsilon_{im_1})$ for every state of the perturber included in the calculation and for every internuclear distance $R$ of interest. However, these calculations can be efficiently parallelized.
			
			On the contrary to the method discussed here, \textcite{Khuskivadze} do not use the \rmat to match the wave functions on the sphere $\mathcal{S}$. Instead, they numerically integrated solutions $q_l(r)$ defined by the single-channel version of \eqab{eq:fexpand}. Another approach was presented by \textcite{Hamilton-PhD} who parametrized the solution in the inner region by the phase shifts and introduced the model relation between the local electron momentum and the total energy of the bound state.
	\section{Numerical aspects of the calculations}
		\label{sec:numeri}
		As explained in Section \ref{sec:innerreg}, the two-electron basis set \eqref{eq:2ebasis} consists of the closed and open subsets
		\begin{subequations}
			\begin{eqnarray}
				y_{n_1l_1n_2l_2}^{(L)\mathrm{C}}(\mathbf{r}_1,\mathbf{r}_2)&=&\frac{C_\gamma}{r_1r_2}\hat{A}\left[\vphantom{\mathcal{Y}_{l_1l_2}^{(L)}(\mathbf{\Omega}_1,\mathbf{\Omega}_2)}u_{n_1l_1}(r_1)u_{n_2l_2}(r_2)\right.\nonumber\\*
				&\times&\left.\mathcal{Y}_{l_1l_2}^{(L)}(\mathbf{\Omega}_1,\mathbf{\Omega}_2)\right],\label{eq:2eclosedsub}\\
				y_{n_1'l_1'n_2'l_2'}^{(L')\mathrm{O}}(\mathbf{r}_1,\mathbf{r}_2)&=&\frac{C_{\gamma'}}{r_1r_2}\hat{A}\left[\vphantom{\mathcal{Y}_{l_1'l_2'}^{(L')}(\mathbf{\Omega}_1,\mathbf{\Omega}_2)}u_{n_1'l_1'}(r_1)B_{\alpha}(r_2)\right.\nonumber\\*
				&\times&\left.\mathcal{Y}_{l_1'l_2'}^{(L')}(\mathbf{\Omega}_1,\mathbf{\Omega}_2)\right]\label{eq:2eopensub},
			\end{eqnarray}
		\end{subequations}
		where $\hat{A}$ provides the spin-adapted antisymmetrization of the expression as shown in \eqab{eq:2ebasis}.
		If the complete set of the closed orbitals $u_{n_1l_1}(r)$ ($n_1=1\dots N_b-1$ for each $l_1$) was used in the open subset \eqref{eq:2eopensub} of the two-particle basis, the closed basis functions \eqref{eq:2eclosedsub} would be redundant, provided that the set of \bspls is complete. However, such two-electron basis set would be very large and the \rmath calculations would not be numerically tractable. In practice, we need only several functions $u_{n_1'l_1'}(r_1)$ to construct the channel functions $\varphi_{im_1}(\mathbf{r}_1)$ defined by \eqab{eq:perchanfuncs}. However, such restricted set of $u_{n_1'l_1'}(r_1)$ combined with the complete set of \bspls in \eqab{eq:2eopensub} leads to an insufficient and asymmetric treatment of the two-electron correlation inside $\mathcal{S}$. We solve this dilemma by introducing the closed part of the basis \eqref{eq:2eclosedsub} in which the one-electron orbitals $u_{n_1l_1}(r_1)$ and $u_{n_2l_2}(r_2)$ span identical sets. The closed part is responsible for the convergence of the correlation interaction inside the sphere $\mathcal{S}$. The open part \eqref{eq:2eopensub} allows one electron to escape the sphere $\mathcal{S}$ and usually requires far fewer one-electron orbitals $u_{n_1'l_1'}(r_1)$.
		
		Apparently, present approach leads to a linear dependence between the open and closed basis function because some of the closed basis elements \eqref{eq:2eclosedsub} can be expressed as proper linear combinations of the open basis functions \eqref{eq:2eopensub}. Some of the molecular \rmath methods \cite{tennyson-rev}, based on different one-particle basis sets than the approach presented here, avoid this dependence by prior orthogonalization of the open radial orbitals $B_\alpha(r_2)$ in \eqab{eq:2eopensub} to the set of $u_{n_2l_2}(r_2)$ selected in \eqab{eq:2eclosedsub}. However, the issue of the overcompleteness of the two-electron basis can be easily solved. First, the singular overlap matrix $\underline{O}$ defined by \eqab{eq:2eovermatels} is diagonalized and its kernel is recognized as the set of eigenvectors corresponding to the zero eigenvalues. The orthonormal projection matrix into the regular subspace is constructed using the non-zero eigenvalues and corresponding eigenvectors \cite{Mayer2002}. Afterwards, the projection of the generalized eigenvalue problem \eqref{eq:geneigen} using that matrix is performed. In addition to elimination of the linear dependence from the two-particle basis set, this procedure reduces the overlap matrix to the identity matrix. The reduction of the generalized eigenvalue problem to the standard one makes its computation more memory efficient because it is not necessary to store the overlap matrix.
		
		Thanks to the block diagonal structure of $\underline{O}$ (see \figref{fig:omatscheme}), its diagonalization can be performed block-wise as well as the projection of $\underline{H}'$ (see \figref{fig:hamscheme}). The dimension of the projected Hamiltonian matrix in the calculations presented here is approximately 10\% smaller than the dimension of the basis set before the projection. Removal of the redundant two-electron space is numerically well defined procedure. In our calculations, all the eigenvalues of the subspace to be removed were at least five orders of magnitude smaller than the rest of the spectrum.
		
		The angular component of each basis function \eqref{eq:2ebasis} is characterized by the quantum numbers $l_1$, $l_2$, $l_1'$, $l_2'$, $L$ and $L'$ and their upper limits influence the convergence of the calculation. The range of $l_2'$ in the set of the open functions \eqref{eq:2eopensub} must be large enough to achieve a sufficient angular resolution of the wave function on the sphere $\mathcal{S}$ and our tests showed that partial waves up to $l_2'=4$ is necessary to achieve converged results in case of the hydrogen molecule. The range of $l_1$ and $l_2$ available in the subset of the closed functions \eqref{eq:2eclosedsub} must be large enough to allow a correct representation of the complicated structure of the correlation cusps of the two-electron wave function in the inner region. The functions with $l_{1,2}\leq4$ were used in the calculations discussed below as a result of the requirement that the two-particle basis set must reproduce the electron affinity of the hydrogen atom with accuracy better than 1\thinspace meV. The requirement on $l_1'$ in the open two-particle basis functions is that a sufficient angular basis for the expansion of few lowest wave functions $\varphi_{im}(\mathbf{r}_1)$ in terms of the spherical harmonics must be provided. $l_1'\leq1$ was sufficient in the calculations presented below.
		
		No constraint is applied the total angular momentum $L$. Therefore, for given pair $l_1$, $l_2$, the two-particle functions $y_\gamma^{(L)}$ for each $\left|l_1-l_2\right|\leq L\leq l_1+l_2$ are included in the basis set.
		
		The radial components of the two-electron basis functions are characterized by indices $n_1$, $n_2$, $n_1'$ and $n_2'$. The set of 90 \bspls of the sixth order was used to obtain the results presented below and the radius of the sphere $\mathcal{S}$ was set to $r_0=10\thinspace$ a.u. The range of $n_1$ in the open subset is usually relatively small ($n_1'\leq2$ in the calculations discussed in this work), since usually only few lowest channels $\varphi_{im_1}(\mathbf{r}_1)$ are involved in the calculations. Even those are very weakly perturbed atomic states of the perturber. Therefore, their expansions in terms of $u_{n_1l_1}(r_1)$ have one dominant term for each $l_1$ and few small corrections. Only the ground state of the perturber influenced by the core \corb was included in the calculations presented here. Considerably larger sets of the radial orbitals are necessary in the closed subsets of $y_\gamma^{(L)}$ to treat the correlation effects of the electron-electron interaction. The converged results presented below were obtained for $n_1,n_2=1\dots20$. All the angular coupling coefficients were calculated using the routines by \textcite{Wei1999} that provide accurate values for higher angular momenta.
		
		A partially parallelized computer implementation of the method discussed in Section \ref{sec:rmatmethod} was developed and executed on the computer cluster with 60 central processing unit (CPU) cores available for this project. Evaluation of every diagonal $L$-block of the modified Hamiltonian matrix $\underline{H}'$ (see \figref{fig:hamscheme}) in the basis set discussed above required approximately 10\thinspace GB of the computer memory, predominately due to the transformation of the matrix elements of the operator $1/r_{12}$ from the \bspl basis to the two-electron basis \eqref{eq:2ebasis}. The amount of the computer memory required for the other steps of the construction of the diagonal blocks of $\underline{H}'$ and $\underline{O}$ is negligible. The total CPU time necessary to calculate all the $R$-independent part of $\underline{H}'$ was approximately 300\thinspace seconds on the Intel Xeon CPU (Ivy Bridge generation). Another step of the calculation that requires considerable computer resources, is the complete diagonalization of $\underline{H}'$ after its projection on the regular subspace. Total dimension of $\underline{H}'$ in the basis set discussed above was 13122, the projection reduced it to 12770. The diagonalization was performed using the parallelized routine dsyev from the program library LAPACK. The diagonalization required 1.2\thinspace GB of the computer memory and 1300\thinspace seconds of the CPU time.
		
\begin{figure}
			\includegraphics{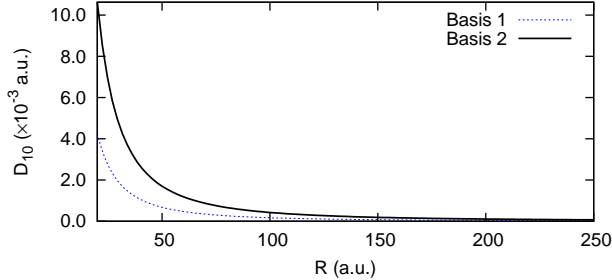}
			\caption{Dipole moment $D_{10}$ of the ground state $\varphi_{10}(\mathbf{r}_1)$ of hydrogen calculated using \eqab{eq:dipexplic} as function of the internuclear separation $R$. The blue dotted curve denoted as Basis 1 was calculated using the same set of the closed orbitals as was used to calculate the energies of the two-electron system. The solid black line denoted as Basis 2 was calculated using the extended set of lowest three $s$-orbitals, lowest two $p$-orbitals and the lowest $d$-orbital.\label{fig:dipdep}}
		\end{figure}

		The potential of the perturber polarized by the atomic core \corb \eqref{eq:outerpoten} is neglected in the outer region. To justify this approximation numerically for the hydrogen perturber, we evaluated the dipole term \eqab{eq:dippoten}, generated by the ground state $\varphi_{10}(\mathbf{r}_1)$, that is the leading term of this potential. Dependence of the corresponding dipole moment $D_{10}$ on the internuclear separation $R$ obtained using \eqab{eq:dipexplic} is shown in \figref{fig:dipdep}. The curve calculated using the same set of the closed orbitals $u_{n_1'l_1'}(r_1)$ (see \eqab{eq:2eopensub}) as discussed above ($n_1'\leq2$, $l_1'\leq1$) is denoted as Basis 1. To study the convergence of the calculated dipole moment, we performed one more calculation of $D_{10}(R)$, where the set of the closed orbitals was extended to three lowest $s$-orbitals, two $p$-orbitals and one $d$-orbital. These results are also plotted in \figref{fig:dipdep} and corresponding curve is denoted as Basis 2. As can be seen, $D_{10}(R)$ calculated in the smaller basis set is considerably lower compared to the one calculated using the more extensive set of orbitals 2, although the two-electron PECs calculated using the smaller basis set 1 are converged (as will be discussed in Section \ref{sec:secresults}). This suggests a higher sensitivity of $D_{10}(R)$ to the set of closed functions $u_{n_1'l_1'}(r_1)$ employed in the open part of the two-electron basis set \eqref{eq:2eopensub}. Our tests (not published here) showed that the curve marked as Basis 2 in \figref{fig:dipdep} does not considerably change with further extension of the set of employed $u_{n_1'l_1'}(r_1)$. When any of the calculated dipole moments is substituted into \eqab{eq:dippoten} for the dipole potential, the value of $U_{D10}(\mathbf{r}_2)$ anywhere in the outer region is more than two orders of magnitude smaller than the Coulomb potential of the core \corb. Consequently, the potential \eqref{eq:outerpoten} can be neglected outside the sphere $\mathcal{S}$.
		
		It is interesting to analyze the relation between the dependence of $D_{10}$ on the internuclear distance $R$ and the static dipole polarizability of the perturber $\alpha_d$.
		As long as $R$ is large enough and the electric field generated by the atomic core \corb can be within a good approximation considered constant throughout the region of the perturber, the dipole moment $D_{10}$ depends on $R$ as $D_{10}(R)=\alpha_d/R^2$ \cite{Khuskivadze,bottcher,Henriet1984}. Comparison of this dependence with $D_{10}$ calculated using \eqab{eq:dipexplic} for different values of $R$ allows us to evaluate the range of $R$ where this approximation of the constant electric field inside $\mathcal{S}$ is valid.
		
		The values of $D_{10}$ obtained from \eqab{eq:dipexplic} using the smaller set of the closed orbitals (Basis 1 in \figref{fig:dipdep}) can be very accurately fitted to
		\begin{equation}
			D_{10}(R)=a/R^2,\label{eq:dipfit}
		\end{equation}
		where $a=1.696$\thinspace a.u. This suggests that the variation of the electric field inside $\mathcal{S}$ does not influence the polarization of the perturber by the core \corb. The low value of $a$ compared to the static dipole polarizability of the hydrogen atom $\alpha_d=4.5$\thinspace a.u. is the consequence of the insufficient set of the closed orbitals $u_{n_1'l_1'}(r_1)$ for the representation of the dipole moment mentioned above.
		
		The curve denoted as Basis 2 in \figref{fig:dipdep}, calculated using more extensive set of the closed orbitals, also can be very accurately fitted to \eqab{eq:dipfit}, where $a=4.25$\thinspace a.u. This is much closer to the correct value of $\alpha_d$. The reciprocal quadratic dependence of $D_{10}$ on the internuclear separation also shows that the presence of the atomic core \corb does not considerably affect the static dipole polarizability of the perturber throughout the studied range of $R$.

		The computationally most demanding part of the outer region treatment is the numerical evaluation of the matrix elements of the \gfn and its radial derivative \eqref{eq:gmatrices}. The three-dimensional integration (over the angular coordinates $\theta$, $\theta''$ and $\phi-\phi''$) \cite{Khuskivadze} is performed numerically using the Gauss-Legendre quadrature in each dimension. The convergence tests showed that the results presented below are converged for 20 abscissas in each dimension. The $1/\left|\mathbf{r}-\mathbf{r}''\right|$ singularity is removed from the \gfn and separately integrated analytically as it is explained in the Appendix of the paper \cite{Khuskivadze}. The Whittaker functions along with their derivatives necessary to evaluate the Coulomb \gfn \cite{Hostler1963} are calculated using the subroutines by \textcite{Thompson1985}. Note that \textcite{Hamilton-PhD} employed a Taylor expansion of the \gfn on the sphere $\mathcal{S}$ assuming that $r_0\ll R$, since the authors focused on the limit of large internuclear separations. That approach requires less evaluations of the complicated special functions. Moreover, this work is oriented more towards the situations, when $r_0$ and $R$ are comparable.
		
		The matrix $\underline{M}$ calculated for fixed $R$, as a function of the energy $E$, possesses two kinds of poles. These are \rmath poles (see \eqab{eq:rmatpolex}) and the poles of the \gfn at the bound-state energies of the Rydberg atom in the absence of the perturber. Since all these are known in advance, the energy interval of our interest was split into the subintervals defined by these poles. The situations, where the bound-state of whole the diatomic system has the energy like the Rydberg atom without the perturber, are not the main focus of this computational method and it is less accurate there. The energy grid between each two poles of $\underline{M}$ for the calculations presented below consisted of 3000 energy points and the evaluation of $\underline{M}$ (its dimension is 5) required 600\thinspace seconds of the CPU time.
	\section{Results}
		\label{sec:secresults}
		As a simple demonstration of the presented computational method, we applied it to H$_2$ for calculations of the PECs of its excited $^1\Sigma$ states. The most simple system was selected intentionally, as its treatment does not require any additional parameters that are necessary for other first-row elements and it is free of all the model-related complications. Both model potentials $V_A(r)$ and $V_B(r')$ reduce to pure Coulomb potentials everywhere in the space and the \gfn does not include any quantum-defect corrections \cite{Davydkin,Khuskivadze,Hamilton-PhD}. Moreover, the results obtained using the method presented here can be compared with those obtained using simple contact-potential models published previously. The technical details of the calculations are discussed above in Section \ref{sec:numeri}.
		
		The PECs for the internuclear separations from 20\thinspace a.u. to 420\thinspace a.u. are presented and they include the nuclear repulsion term $1/R$ that is not part of the electronic Hamiltonian \eqref{eq:totham}. In order to keep the graphs easier to read, the energy -0.5\thinspace a.u. of H($1s$) was subtracted from all the total energies. In the limit $R\to\infty$, every bound-state PEC converges to an energy of the non-interacting system of the perturber in its ground state and the atomic Rydberg state of the other atom $\mathrm{H}(1s)+\mathrm{H}(nl)$. These energies as well as corresponding groups of the PECs will be denoted by $n$ in the rest of this section.
		
\begin{figure}
			\includegraphics{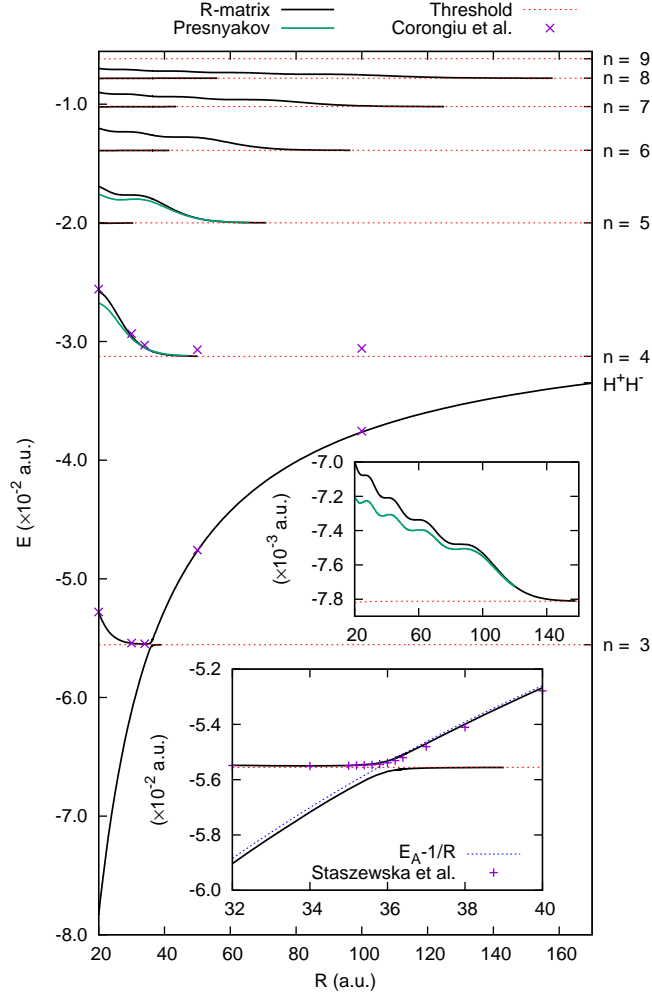}
			\caption{Potential energy curves of the excited $^1\Sigma$ states of H$_2$. The nuclear repulsion is included and the energy -0.5\thinspace a.u. of H$(1s)$ was subtracted. The \rmath results are plotted with the solid black line, the results of the contact model \cite{Presnyakov} are represented by the solid green line. The dotted red lines are the poles of the \gfn $\epsilon_{10}-1/(2n^2)+1/R$. The blue dotted line in the inset corresponds to $E_A-1/R$. The points denoted by $\times$ are the energies calculated by \textcite{Corongiu}, the points denoted by $+$ in the lower inset are the energies calculated by \textcite{Staszewska}.\label{fig:lowstates}}
		\end{figure}

		For finite internuclear distances, the Coulomb \gfn has poles at energies $\epsilon_{10}(R)-1/(2n^2)$, where the channel energy $\epsilon_{10}(R)$ is defined by \eqab{eq:perchanfuncs} and $n\in\mathbb{N}$. Those energies would correspond to eigenstates of a system in which the Rydberg electron does not interact with the neutral atom with the center \cora. They will be referred to as the non-interacting Rydberg thresholds (NIRTs) in the rest of this section. The energy of the excited channel $\epsilon_{20}(R)$ is so high that all the corresponding NIRTs are above the ionization threshold. Therefore, the excited channels were not included in the outer region treatment. \figref{fig:lowstates} demonstrates (the red dotted curves) that NIRTs exhibit only very weak dependence on $R$ over the studied interval
		because the ground state of the perturber $\varphi_{10}(\mathbf{r}_1)$ is only very weakly perturbed by the atomic core \corb. The first-order perturbation theory yields $\epsilon_{10}(R)=\varepsilon_{10}-1/R$ and addition of the nuclear repulsion cancels the dependence on the internuclear distance. The good agreement of this result with the non-perturbative calculations presented in \figref{fig:lowstates} means that the variation of $V_B$ over the spatial volume, where the wave function of the valence electron is located, does not play any considerable role. This is understandable taking into account that the wave function of H$(1s)$ is very compact.
		\begin{figure}
			\includegraphics{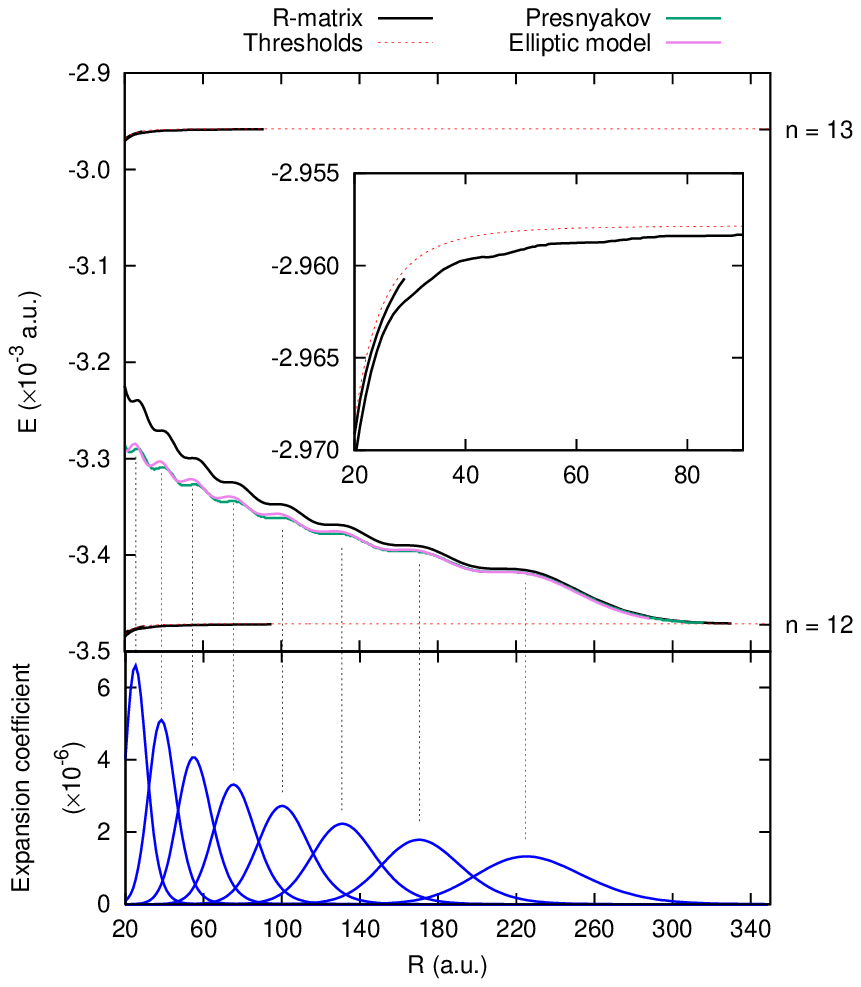}
			\caption{Top: PECs of the excited $^1\Sigma$ states of H$_2$. The curves have the same meaning like in \figref{fig:lowstates}, results between NIRTs corresponding to $n=12$ and $n=13$ are presented. Bottom: Electron density of the spheroidal eigenstates $\left|\tau_{i0n}(\mathbf{r}_{\text{pert}})\right|^2$ of H$(n=12)$ located in one focus \cite{Coulson} evaluated at the position of the other focus as a function of the distance between the foci.\label{fig:curdetail}}
		\end{figure}
		As can be seen in the inset of \figref{fig:curdetail}, the higher-order perturbation becomes more pronounced at the internuclear separations $R\lesssim40$\thinspace a.u.
		
		The PEC on the bottom of \figref{fig:lowstates} is the ion-pair curve. The lower inset of that figure shows that for $R\lesssim40$\thinspace a.u., the ion-pair PEC is slightly diverted from its asymptotic form $-0.5-E_A-1/R$, where $E_A$ is the electron affinity of H. This is another consequence of the increased polarization of the perurber by the core \corb at smaller $R$. Note that the position and shape of the avoided crossing of the ion-pair PEC with the $4^1\Sigma_u^+$ at $R\approx36$\thinspace a.u. is in excellent agreement with the previously published results of the \abini quantum chemical calculations by \textcite{Staszewska}. Moreover, the results of \textcite{Corongiu} for higher energies are also in good agreement with the findings presented here. The accuracy of the ion-pair PEC is sensitive predominately to the quality of the representation of the two-electron interaction in the inner region, since the probability density of the Rydberg electron in this state is well localized around the perturber. It is the convergence of this ion-pair PEC in our calculations that requires the set of the closed two-electron basis functions larger than the subset of the open functions.
		
\begin{figure}
			\includegraphics{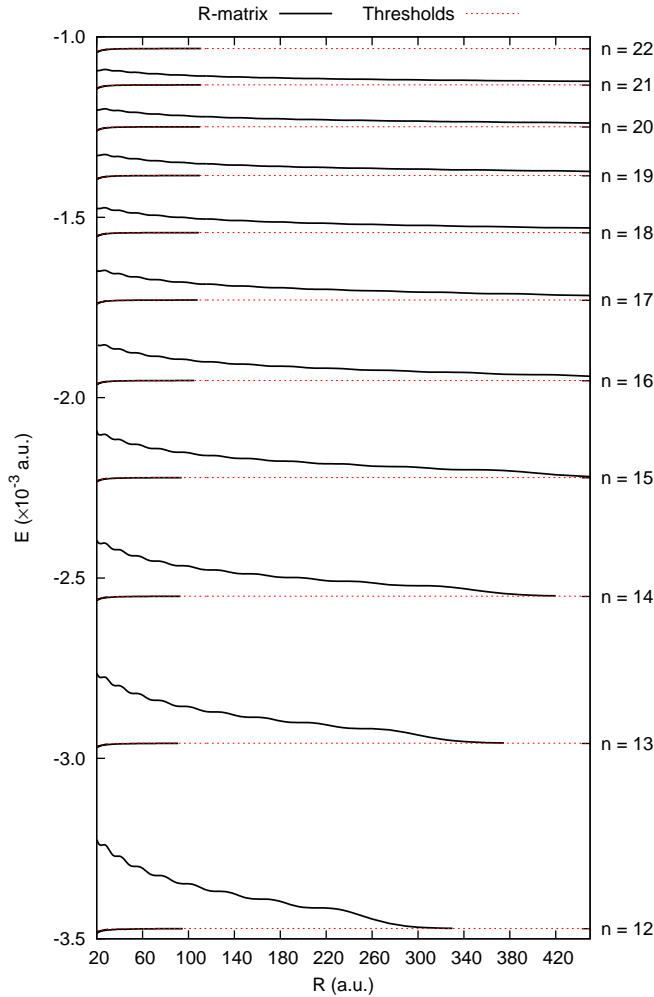}
			\caption{PECs of the excited $^1\Sigma$ states of H$_2$. The interval of the asymptotic energies of $\mathrm{H}(1s)+\mathrm{H}(nl)$ with $n=12\dots22$ is presented. The colors and the notation has the same meaning as in \figref{fig:lowstates}.\label{fig:assem}}
		\end{figure}

		As can be seen in \figref{fig:assem}, the PECs 
		(other than the ion-pair curve) can be divided into two categories: The PECs of the first type follow the shape of the $1\sigma_g$ PEC of H$_2^+$ at large internuclear separations and they are located at energies very close below the NIRTs (see also the inset of \figref{fig:curdetail}). These PECs represent the Rydberg states weakly affected by the neutral perturber. For the energy interval of the interest in this work these PECs coincide with the associated NIRTs at $R\gtrsim120\thinspace$a.u. The PECs of the second type are more interesting. They are located at energies between the pairs of subsequent NIRTs at smaller internuclear separations and each of them decreases with increasing $R$ until it coincides with the lower NIRT.
		
		PECs of both types show an oscillatory structure that is better pronounced in the second-type PECs. These undulations can be understood in terms of the simplified model developed by \textcite{Granger} to predict the trilobite-like states of Rb$_2$ \cite{Greene-prl}. In their single-electron approach the perturber is represented by Fermi's contact pseudopotential and the Rydberg center by the pure Coulomb potential (the non-hydrogen character of Rb is neglected). The eigenstates of this model system are expressed as linear combinations of such eigenfunctions $\tau_{nmi}(\mathbf{r})$ of the hydrogen-atom Hamiltonian that are separable in the prolate spheroidal coordinates \cite{Coulson} with the foci located on the atomic cores. The quantum number $m$ is set to zero, since only the $\Sigma$ states are considered. The quantum number $i$ denotes different degenerate eigenstates with the same $n$. \textcite{Granger} applied the first-order perturbation theory to calculate the energies of the model diatomic system as
		\begin{equation}
			E_n(R)=-\frac{1}{2n^2}+2\pi A\left[k(R)\right]\sum_{i}\left|\tau_{i0n}(\mathbf{r}_{\text{pert}})\right|^2,\label{eq:greenepert}
		\end{equation}
		where $n$ is the principal quantum number of the atomic Rydberg electron without the perturber, $A\left[k(R)\right]$ is the electron-perturber scattering length and the values of the atomic spheroidal eigenstates are taken at the position of the perturber. \textcite{Granger} showed that in case of Rb$_2$ every minimum in the oscillations of the PEC corresponds to such values of $R$, where the sum in \eqab{eq:greenepert} is dominated by one spheroidal eigenstate of the Rydberg electron. The values of $\left|\tau_{i0n}(\mathbf{r}_{\text{pert}})\right|^2$ for $n=12$ are plotted in the lower part of \figref{fig:curdetail}. As can be clearly seen, the structures in the PECs calculated using the method introduced in this work obviously correspond to the peaks of the spheroidal eigenstates of the hydrogen atom and they are of the same nature like those in Refs. \cite{Greene-prl,Granger}.
		
		More specifically, the peaks of the spheroidal eigenstates correspond to the local maxima of the oscillations in the PECs. This behavior is different from the case of Rb$_2$ discussed by \textcite{Granger}, where the positions of the peaks of the spheroidal atomic eigenstates coincide with the minima of the oscillations. An explanation is quite simple: The scattering length $A(k)$ in \eqab{eq:greenepert} is negative for hydrogen and positive for rubidium \cite{Schwartz,Bahrim2000}.
		
		It is interesting to compare the results of our two-electron \rmath calculations with those obtained by the application of the approach developed by \textcite{Greene-prl} and \textcite{Granger} to H$_2$. The generalized energy-dependent scattering length for the interaction of the free electron with the neutral atom is defined as $A(k)=-\tan\delta(k)/k$, where $\delta(k)$ is the $s$-wave scattering phase shift. The effective range theory modified for the presence of the polarization potential \cite{OMalley1961} provides the low-energy expansion
		\begin{equation}
			-\frac{1}{A(k)}=-\frac{1}{A_0}+\frac{\pi\alpha_d}{3A_0^2}k+\frac{4\alpha_d}{3A_0}k^2\ln\left(\frac{\sqrt{\alpha_d}k}{4}\right)+\mathcal{O}(k^2),\label{eq:mert}
		\end{equation}
		where $A_0=5.95$ is the singlet zero-energy scattering length \cite{Schwartz} and $\alpha_d=4.5$\thinspace a.u. is the static dipole polarizability of the hydrogen atom. All the terms where $k$ appears in the second and higher order will be neglected here. Following \textcite{Greene-prl}, the local electron momentum $k$ can be for given internuclear separation $R$ calculated as
		\begin{equation}
			k(R)=\sqrt{2\left(\frac{1}{R}-\frac{1}{2n^2}\right)}.\label{eq:localmom}
		\end{equation}
		Substitution of \eqab{eq:localmom} and \eqab{eq:mert} into \eqab{eq:greenepert} yields the energies of the model diatomic system. The PEC calculated for $n=12$ is plotted in \figref{fig:curdetail}. It is in encouraging qualitative agreement with the two-electron \rmath calculations, including the details of the oscillatory structure. It is not surprising that the agreement is excellent at larger internuclear separations ($R\gtrsim100$\thinspace a.u.) and that the results of the one-electron model increasingly divert from the PEC calculated using the two-electron \rmath approach with decreasing $R$. Comparable agreement between the Fermi's model and two-electron \rmath calculations was observed for all the energies and internuclear separations $R$ explored in this paper. This is a consequence of the finite size of the perturber that plays more important role at smaller internuclear separations, while the representation by the Fermi's contact pseudopotential reduces the interaction with the perturber to a single spatial point.
		
		Fermi's pseudopotential \cite{fermi1934}
		\begin{equation}
			V_{Fk}(\mathbf{r})=2\pi A\left[k(R)\right]\delta^3(\mathbf{r}-\mathbf{r}_{\text{pert}}),
		\end{equation}
		previously applied to Rb$_2$ \cite{Greene-prl,Granger}, is parametrized to treat a collision of a free electron with a neutral target. While this model holds very well for large internuclear separations ($R\sim10^3 - 10^5$\thinspace a.u. \cite{Greene-prl}), some of its assumptions cease to be accurate at smaller distances $R$ explored in this paper. For instance, the dipole moment of the neutral atom gained in the presence of the potential $V_B$ becomes relevant inside the sphere $\mathcal{S}$, although we neglect it in the outer region. Moreover, variation of $V_B$ inside the sphere $\mathcal{S}$ is larger for smaller $R$ and the validity of the free-electron scattering model becomes questionable. We attempt to account for this deficiency by adding the third, $k^2\ln k$, term in the expansion \eqref{eq:mert} that appears to be not included in the original works \cite{Greene-prl,Granger}. This term becomes important because larger variation of the potential $V_B$ around the perturber provides the scattered electron with higher local momenta via \eqab{eq:localmom}. In fact, we also carried out tests (not published here) with only first two terms on the right-hand side of \eqab{eq:mert} and the agreement of such Fermi's model with the two-particle \rmath calculations was very poor.
		
		Another simple model worth mentioning here was developed by \textcite{Presnyakov}. It is based on a different Fermi's pseudopotential \cite{blatt-book} and \textcite{Presnyakov} utilized it to study broadening of the excited levels of atoms in an alkali-metal atmosphere. The author solved the \sche non-perturbatively using the \gfn \cite{Hostler1963} and the PECs of the excited bound states $E(R)$ were calculated as the solutions of the transcendental equation
		\begin{eqnarray}
			-\frac{1}{A_0}&=&\frac{2\Gamma(1-\kappa)}{\kappa}\left\{\left[-\frac{1}{4}+\frac{\kappa}{z}\right]\mathcal{W}_{\kappa,1/2}(z)\mathcal{M}_{\kappa,1/2}(z)\right.\nonumber\\
			&+&\left.\mathcal{W}'_{\kappa,1/2}(z)\mathcal{M}'_{\kappa,1/2}(z)\right\},\label{eq:presnyakov}
		\end{eqnarray}
		where $\kappa=\sqrt{-2E(R)}$, $z=2R/\kappa$, $\mathcal{W}_{\kappa,1/2}(z)$ and $\mathcal{M}_{\kappa,1/2}(z)$ are Whittaker functions \cite{Hostler} and the prime denotes the derivative with respect to the argument. Therefore, the computational method is not based on the perturbation theory. We solved \eqab{eq:presnyakov} numerically to compare the results of this model with the simple models discussed above \cite{Granger,Greene-prl} as well as with our two-electron \rmath results. The PECs obtained by solving \eqab{eq:presnyakov} are in excellent agreement with the results of the first-order perturbation calculations \cite{Granger,Greene-prl}, as can be seen in \figref{fig:curdetail} (the agreement is similar throughout all the energy range shown). Such a good agreement indicates that the first-order perturbation theory is very accurate when applied to the states separable in the spheroidal coordinates. Comparison between the solutions of \eqab{eq:presnyakov} and the PECs obtained using our two-electron \rmath approach for the lower energies is plotted in \figref{fig:lowstates}. Both calculations are in very good qualitative agreement. They agree very well quantitatively at larger internuclear separations and as $R$ decreases, the energies of the one-electron model are systematically slightly lower than those obtained from the \rmath calculations.
	\section{Conclusions}
		\label{sec:coclusions}
		A method for study of long-range Rydberg states of diatomic molecules was introduced in this paper. The method can be viewed as a straightforward extension of the pure one-electron approach of \textcite{Khuskivadze}. In the present work, the two-electron \rmath method is employed to calculate the wave function in the region surrounding one of the atomic cores. The wave function smoothly connects to a bound-state solution dictated by a long-range Coulomb field of the second atomic center. The approach presented here does not account for a situation in which both electrons leave the first atomic center and thus the ionization energy of the first core defines the upper energy limit of this approach.
		
		In comparison to the method of \textcite{Khuskivadze}, the present approach brings a few simplifications and some complications. One of the simplifications is the core potential $V_A$ that can be made independent of the total spin state of the system as well as of the total angular momentum with respect to the first atomic core. These properties allow calculations for much shorter internuclear distances $R$ than those explored by \textcite{Khuskivadze}. The shorter internuclear separations $R$ require higher partial waves with respect to the first center and this requirement raises the demands on the accurate representation of the neutral perturber (center \cora) that was considered different for the partial waves $l=0$ and $l=1$ and every total spin state in \textcite{Khuskivadze}. On the other hand, the two-electron method presented here has to deal with the correlated movement of the electrons in the core potential $V_A$ inside a sphere surrounding the atomic core \cora. Such complication may also become a gain in situations, where the energies of the molecular Rydberg states associated with the excited states of the perturber are below the ionization threshold of the whole system (for example, Rb$_2$).
		
		The motivation for the development of the presented method is to study the long-range Rydberg states of the alkali-metal diatomics at the internuclear separations $R\approx20-400$\thinspace a.u., corresponding to experiments \cite{Greene2006-prl,Vadla,Bellos}. This work aims to complement previous studies \cite{Greene2006-prl,Bellos} and test the approximation of the contact potential at smaller internuclear distances utilized there to explain the experimental spectra. In order to test the basic mechanisms of the method we decided to test it on the simplest of molecules, H$_2$. Present results accurately reproduce previous \abini calculations \cite{Corongiu,Staszewska} for lower states ($n=3,4$) and they predict yet undocumented oscillatory character of higher anti-bonding states $\mathrm{H}(1s)+\mathrm{H}(nl)$. While these oscillations are also well reproduced by the two contact-potential models tested in this paper, the absolute energies predicted by the contact-potential models are systematically lower than the more accurate two-particle calculations. Present numerical analysis in the spheroidal coordinates reveals that the physical origin of these oscillations is the same as was explained by \textcite{Granger} for calculated trilobite states, i.e. the oscillations correspond to the peaks of the dominant spheroidal eigenstates.
		
		Our internal numerical tests indicate that the accuracy of the presented energies should be better than 1\thinspace meV. Therefore, we believe that the present method should bring similar level of accuracy to the characterization of the long-range molecular Rydberg states of the alkali-metal molecules. Such studies are planned for the near future.
	\begin{acknowledgments}
		We are thankful to Chris Greene for the initial impulse for this work. We also acknowledge the support of the Grant Agency of the Czech Republic (Grant No. P208/14-15989P).
	\end{acknowledgments}
	\appendix
	\section{Matrix elements of the off-center Coulomb potential}
		\label{sec:append}
		Since the matrix elements of $V_B$ inside the sphere for the first and second electron are equal, only one of them is derived here. The  multipole expansion \eqref{eq:polexpand} can be simplified for $r_1<R$, $\mathbf{R}=(0,0,-R)$ as follows:
		\begin{equation}
			\frac{1}{\left|\mathbf{r}_1-\mathbf{R}\right|}=\sum\limits_{\lambda=0}^{\infty}(-1)^\lambda\sqrt{\frac{4\pi}{2\lambda+1}}\frac{r_1^\lambda}{R^{\lambda+1}}Y_{\lambda0}(\mathbf{\Omega}_1).
			\label{eq:offpolex}
		\end{equation}
		It can be seen from this equation that the matrix elements in the basis \eqref{eq:2ebasis} can be expressed as a sum of the products of the radial and angular factors.
		
		The angular part can be easily calculated using the algebra of the two-particle irreducible spherical tensors \cite{Sobelman1972,edmonds} as
		\begin{widetext}
			\begin{eqnarray}
				\chi_{l_1l_2l_{1}'l_{2}'\lambda}^{(LL')}&=&(-1)^\lambda\sqrt{\frac{4\pi}{2\lambda+1}}\Braket{\mathcal{Y}_{l_1l_2}^{(LM)}(\mathbf{\Omega}_1,\mathbf{\Omega}_2)|Y_{\lambda0}(\mathbf{\Omega}_1)|\mathcal{Y}_{l_1'l_2'}^{(L'M)}(\mathbf{\Omega}_1,\mathbf{\Omega}_2)}=(-1)^{L-M+L'+l_2}\nonumber\\*
				&\times&\sqrt{(2l_1+1)(2l_1'+1)(2L+1)(2L'+1)}
				\begin{Bmatrix}
					l_1&L&l_2\\
					L'&l_1'&\lambda
				\end{Bmatrix}
				\begin{pmatrix}
					l_1&\lambda&L'\\
					0&0&0
				\end{pmatrix}
				\begin{pmatrix}
					L&\lambda&L'\\
					-M&0&M
				\end{pmatrix}\delta_{l_2l_2'}.
			\end{eqnarray}
		\end{widetext}
		To keep the notation more simple, $M$ is not listed as an argument of $\chi$, since it is a parameter of the calculation fixed at the beginning.
		It is obvious from the 3-j and 6-j symbols that these angular factors are non-zero only if all their triangularity conditions are satisfied.
		This means that for given $l_1,l_2,l_1',l_2',L,L'$, the sum in \eqab{eq:offpolex} is restricted by $\lambda_{\min}\leq\lambda\leq\lambda_{\max}$:
		\begin{subequations}
			\label{eq:selrules}
			\begin{eqnarray}
				\lambda_{\min}&=&\max(\left|L-L'\right|,\left|l_1-L'\right|,\left|l_1-l_1'\right|)\\
				\lambda_{\max}&=&\min(L+L',l_1+L',l_1+l_1')
			\end{eqnarray}		
		\end{subequations}
		
		The multipole radial factors
		\begin{equation}
			\rho_{n_1l_1n_2l_2\lambda}=\int\limits_{0}^{r_0}dr
			\begin{Bmatrix}
				u_{n_1l_1}(r)\\
				f_{n_1l_1}(r)
			\end{Bmatrix}
			r^\lambda
			\begin{Bmatrix}
				u_{n_2l_2}(r)\\
				f_{n_2l_2}(r)
			\end{Bmatrix}
		\end{equation}
		include all four one-electron terms. These integrals are first numerically evaluated in the \bspl basis and then the closed part is transformed using the coefficients $b_{\alpha nl}$ in \eqab{eq:bexcoefs}. Note that none of the angular nor radial factors depend on the internuclear separation $R$. Using all these factors, the matrix element in the two-electron basis \eqref{eq:2ebasis} can be expressed as
		\begin{eqnarray}
			\Braket{y_\gamma^{(L)}|\frac{-1}{\left|\mathbf{r}_1-\mathbf{R}\right|}|y_{\gamma'}^{(L')}}&=C_\gamma C_{\gamma'}&\sum\limits_{\lambda=\lambda_{\min}}^{\lambda_{\max}}\frac{-1}{R^{\lambda+1}}\nonumber\\*
			\times\left[\rho_{n_1l_1n_1'l_1'\lambda}\vphantom{\chi_{l_1l_2l_1'l_2'\lambda}^{(LL')}}\right.\;&\rho_{n_2l_2n_2'l_2'0}&\;\chi_{l_1l_2l_1'l_2'\lambda}^{(LL')}\nonumber\\*
			+\Pi\rho_{n_2l_2n_1'l_1'\lambda}\;&\rho_{n_1l_1n_2'l_2'0}&\;\chi_{l_2l_1l_1'l_2'\lambda}^{(LL')}\nonumber\\*
			+\Pi'\rho_{n_1l_1n_2'l_2'\lambda}\;&\rho_{n_2l_2n_1'l_1'0}&\;\chi_{l_1l_2l_2'l_1'\lambda}^{(LL')}\nonumber\\*
			+\Pi\Pi'\rho_{n_2l_2n_2'l_2'\lambda}\;&\rho_{n_1l_1n_1'l_1'0}&\;\left.\chi_{l_2l_1l_2'l_1'\lambda}^{(LL')}\right],
			\label{eq:offpotel}
		\end{eqnarray}
		where $\Pi=(-1)^{l_1+l_2-L+S}$ and $\Pi'=(-1)^{l_1'+l_2'-L'+S}$. Note that the factors $\rho$ with $\lambda=0$ are the overlap integrals between the radial orbitals. Their presence is a direct consequence of the one-electron nature of $V_B$.
		
	\section{\textsl{R}-matrix surface amplitudes in the uncoupled channels}
		\label{sec:amplider}
		Following \textcite{Robicheaux1991,Aymar1996}, let us start with the definition of the surface harmonics $\Phi_{\bar{j}}(\mathbf{r}_1,\mathbf{\Omega}_2)$ so that the general solutions of the \sche on the surface $\mathcal{S}$ can be expressed as 
		\begin{equation}
			\left.\Psi_\beta(\mathbf{r}_1,\mathbf{r}_2)\right|_{r_2=r_0}=\sum\limits_{\bar{j}}\Phi_{\bar{j}}(\mathbf{r}_1,\mathbf{\Omega}_2)\frac{1}{r_0}q_{\bar{j}\beta}(r_0).
		\end{equation}
		In case of the channel expansion \eqref{eq:chanexpand}, this equation yields
		\begin{equation}
			\Phi_{\bar{j}}(\mathbf{r}_1,\mathbf{\Omega}_2)=\sum\limits_{l_1}\frac{\xi_{il_1m_1}(r_1)}{r_1}Y_{l_1m_1}(\mathbf{\Omega}_1)Y_{l_2(M-m_1)}(\mathbf{\Omega}_2),
			\label{eq:surfharm}
		\end{equation}
		where the wave functions $\varphi_{im_1}(\mathbf{r}_1)$ are expanded in terms of the spherical harmonics as
		\begin{equation}
			\varphi_{im_1}(\mathbf{r}_1)=\sum_{l_1=\left|m_1\right|}^\infty\frac{\xi_{il_1m_1}(r_1)}{r_1}Y_{l_1m_1}(\mathbf{\Omega}_1).\label{eq:perchansphex}
		\end{equation}
		
		Definition of the surface harmonics \eqref{eq:surfharm} and the way how the general solutions on the sphere are expanded in terms of the channels, is to some extent arbitrary. The definition \eqref{eq:chanexpand} and \eqref{eq:surfharm} employed in this paper is motivated by the symmetry of the problem. It allows the wave function that was constructed in the $LS$-coupled representation, to be expressed in terms of the $L$-uncoupled representation. To this end, the spherical harmonics $Y_{l_1m_1}(\mathbf{\Omega}_1)$ and $Y_{l_2(M-m_1)}(\mathbf{\Omega}_2)$ in \eqab{eq:surfharm} are not coupled to form an eigenstate of $\hat{L}^2$.
		
		The target part of $\Phi_{\bar{j}}$ is constructed using the eigenstates $\varphi_{im_1}(\mathbf{r}_1)$ of Hamiltonian $\hat{H}_c$ \eqref{eq:chanham} that includes $V_B(\left|\mathbf{r}_1-\mathbf{R}\right|)$. A formulation of the problem with the atomic radial functions $u_{n_1l_1}(r_1)$ instead of the perturbed radial functions $\xi_{il_1m_1}(r_1)$ in \eqab{eq:surfharm} is also possible and it yields more simple formula for the \rmath amplitudes $w_{\bar{j}k}$. However, it would lead to additional couplings of $u_{nl}(r)$ by $V_B$ in the projection of the \sche on $\Phi_{\bar{j}}$ in the outer region.
		Projection of the basis functions \eqref{eq:2ebasis} onto $\Phi_{\bar{j}}$ on the sphere $\mathcal{S}$ gives
		\begin{eqnarray}
			\Lambda_{\bar{j}\gamma'}^{(L)}&=&\int_{\mathcal{V}}dV_1\int_{\mathcal{S}}d\Omega_2\,r_2^2\Phi_{\bar{j}}(\mathbf{r}_1,\mathbf{\Omega}_2)y_{\gamma'}^{(L)}(\mathbf{r}_1,\mathbf{r}_2)=\nonumber\\*
			&=&C_{\gamma'} r_0f_{n_2'l_2'}(r_0)\delta_{l_2l_2'}(l_1'm_1l_2(M-m_1)|LM)\nonumber\\*
			&\times&\int\limits_{0}^{r_0}dr_1\,u_{n_1'l_1'}(r_1)\xi_{il_1'm_1}(r_1),
			\label{eq:lambdaamps}
		\end{eqnarray}
		where $\mathcal{V}$ is the volume enclosed by the sphere $\mathcal{S}$ and $(l_1'm_1l_2(M-m_1)|LM)$ is the Clebsch-Gordan coefficient \cite{edmonds, Sobelman1972}. The only non-zero elements of matrix $\underline{\Lambda}$ are those where $f_{nl}(r)\equiv B_{N_b}(r)$, since that is the only radial function in the one-particle basis set that does not vanish on the boundary of the inner region.
		
		Large internuclear distances $R$, examined in this paper, cause very weak perturbation of the atomic states $u_{n_1l_1}$. Therefore, the overlap integrals in \eqab{eq:lambdaamps} are close to one for $n_1=i$ and they rapidly decrease with increasing difference of the indices $\left|n_1-i\right|$.
		The \rmath amplitudes in \eqab{eq:rmatpolex} then can be calculated as \cite{Robicheaux1991,Aymar1996}
		\begin{equation}
			w_{\bar{j}k}=\sum\limits_{L,\gamma}c_{\gamma k}^{(L)}\Lambda_{\bar{j}\gamma}^{(L)}.
			\label{eq:surfamps}
		\end{equation}
		Superscript $L$ in the notation of the eigenvector $\mathbf{c}_k$ denotes the block corresponding to the subset of basis functions $\eqref{eq:2ebasis}$ with that value of $L$.
	\section{Dipole moment of the perturber induced by external Coulomb field}
		\label{sec:dipappend}
		Substitution of the multipole expansion \eqref{eq:polexpand} into the first term of \eqab{eq:outerpoten} along with $\varphi_{im_1}(\mathbf{r}_1)$ expressed in terms of the spherical harmonics \eqref{eq:perchansphex} results in
		\begin{eqnarray}
			U_{im_1i'm_1'}(\mathbf{r}_2)&=&\sum_{\substack{\lambda=\left|\mu\right|\\\lambda\neq0}}^{\infty}\frac{Y_{\lambda\mu}(\mathbf{\Omega}_2)}{r_2^{\lambda+1}}\sum\limits_{l_1=\left|m_1\right|}^{\infty}\,\sum\limits_{l_1'=\left|m_1'\right|}^{\infty}\mathcal{Z}_{l_1m_1l_1'm_1'}^{(\lambda)}\nonumber\\*
			&\times&F_{il_1m_1i'l_1'm_1'}^{(\lambda)},\label{eq:outermulti}
		\end{eqnarray}
		where $\mu=m_1'-m_1$. The Coulomb term $\lambda=0$ is missing due to the second term in \eqab{eq:outerpoten}. Factors $\mathcal{Z}$ are the angular integrals of three spherical harmonics \cite{edmonds,Sobelman1972}:
		\begin{eqnarray}
			\mathcal{Z}_{l_1m_1l_1'm_1'}^{(\lambda)}&=&(-1)^{m_1'}\sqrt{\frac{4\pi(2l_1+1)(2l_1'+1)}{2\lambda+1}}\nonumber\\*
			&\times&
			\begin{pmatrix}
				l_1&\lambda&l_1'\\
				0&0&0
			\end{pmatrix}
			\begin{pmatrix}
				l_1&\lambda&l_1'\\
				-m_1&-\mu&m_1'
			\end{pmatrix}\label{eq:dipang}
		\end{eqnarray}
		and factors $F$ are the radial multipole integrals
		\begin{equation}
			F_{il_1m_1i'l_1'm_1'}^{(\lambda)}=\int\limits_{0}^{r_0}dr_1\,\xi_{il_1m_1}(r_1)r_1^\lambda \xi_{i'l_1'm_1'}(r_1).
			\label{eq:radmulti}
		\end{equation}
		Let us evaluate these factors for the permanent dipole term that corresponds to $\lambda=1$, $i=i'$ and $m_1=m_1'$ and let us denote that term $U_{im_1}^D(\mathbf{r}_2)$:
		\begin{equation}
			U_{im_1}^D(\mathbf{r}_2)=\frac{D_{im_1}\cos\theta_2}{r_2^2}
			\label{eq:dippoten}
		\end{equation}
		Properties of the Wigner 3-j symbols in \eqab{eq:dipang} restrict the non-zero angular factors only to the cases where $l_1=l_1'\pm1$ and $\mu=0$. The induced dipole moment is
		\begin{equation}
			D_{im_1}=\sqrt{\frac{3}{4\pi}}\sum\limits_{l_1=\left|m_1\right|}^{\infty}\sum\limits_{\substack{l_1'=l_1\pm1\\l_1'\geq0}}\left[\mathcal{Z}_{l_1m_1l_1'm_1}^{(\lambda=1)}F_{il_1m_1il_1'm_1}^{(\lambda=1)}\right]
			\label{eq:dipexplic}
		\end{equation}
		and the angular factors $\mathcal{Z}$ are
		\begin{eqnarray}
			\mathcal{Z}_{l_1m_1(l_1+1)m_1}^{(\lambda=1)}&=&\sqrt{\frac{4\pi}{3}}\left(\frac{l_1+m_1+1}{2l_1+1}\right)^{1/2}\nonumber\\*
			&\times&\left(\frac{l_1-m_1+1}{2l_1+3}\right)^{1/2}\\
			\mathcal{Z}_{l_1m_1(l_1-1)m_1}^{(\lambda=1)}&=&\left[\frac{4\pi}{3}\frac{(l_1+m_1)(l_1-m_1)}{(2l_1-1)(2l_1+1)}\right]^{1/2}.\phantom{Y}
		\end{eqnarray}
		
		In case of sufficient large internuclear distance $R$, the expansions \eqref{eq:perchansphex} for low $i$ and $m_i$ have typically one dominant term and the remaining terms in the series act as small corrections. This makes the radial integrals \eqref{eq:radmulti} that appear in \eqab{eq:dipexplic}, and consequently, all the permanent dipole moment $D_{im_1}$, small.
		
		The polarization effect due to a positively charged atomic core was theoretically studied by \textcite{bottcher}. Their induced dipole moment was parametrized by the static dipole polarizability, distance of the two atomic cores and by a radial cut-off function to eliminate the unphysical singularity at the origin. Such an approach was later successfully utilized by \textcite{Henriet1984} in calculations of the ground and excited states of the alkali dimers.
%
\end{document}